\documentclass[aps,prl,twocolumn,showpacs,superscriptaddress]{revtex4-1}
\usepackage{graphicx}
\usepackage{dcolumn}
\usepackage{bm}
\usepackage[english]{babel}
\usepackage{amssymb,amsfonts,amsmath,bbold,bm,bbm}
\usepackage{color}
\usepackage{hyperref}
\usepackage{ulem}

\newcommand{\SumInt}[3]{\sum \hspace{-#3} \int\limits_{#1}^{#2}}

\newcommand{\Vsd}{V_{\rm sd}}
\newcommand{\Vg}{V_{\rm c}}
\newcommand{\GQ}{G_{\rm Q}}
\newcommand{\lx}{l_{\rm x}}
\newcommand{\wx}{\Omega_{\rm x}}

\begin{document}

\title{Spin fluctuations in the 0.7-anomaly in quantum point contacts}
\author{Dennis H. Schimmel}
\affiliation{Physics Department, Arnold Sommerfeld Center for Theoretical Physics, and Center for NanoScience,
Ludwig-Maximilians-Universit\"at, Theresienstra\ss e 37, 80333 Munich, Germany}
\author{Benedikt Bruognolo}
\affiliation{Physics Department, Arnold Sommerfeld Center for Theoretical Physics, and Center for NanoScience,
Ludwig-Maximilians-Universit\"at, Theresienstra\ss e 37, 80333 Munich, Germany}
\affiliation{Max-Planck-Institut f\"ur Quantenoptik,
Hans-Kopfermann-Stra\ss e 1, 85748 Garching, Germany}
\author{Jan von Delft}
\affiliation{Physics Department, Arnold Sommerfeld Center for Theoretical Physics, and Center for NanoScience,
Ludwig-Maximilians-Universit\"at, Theresienstra\ss e 37, 80333 Munich, Germany}

\date{\today}

\begin{abstract}
  It has been argued that the $0.7$ anomaly in quantum point contacts
  (QPCs) is due to an enhanced density of states at the top of the
  QPC-barrier (van Hove ridge), which strongly enhances the effects of
  interactions. 
  Here, we analyze their effect on dynamical quantities.  We find that
  they pin the van Hove ridge to the chemical potential when the QPC
  is subopen; cause a temperature dependence for the linear
  conductance that qualitatively agrees with experiment;
  strongly enhance the magnitude of the dynamical spin
  susceptibility; and significantly lengthen the QPC traversal time.
  We conclude that electrons traverse the QPC via a
  slowly fluctuating spin structure of finite spatial extent.
\end{abstract}

\maketitle

Quantum point contacts are narrow, one-dimensional (1D) constrictions
usually patterned in a two-dimensional electron system (2DES) by
applying voltages to local gates. As QPCs are the ultimate building
blocks for controlling nanoscale electron transport, much effort has
been devoted to understand their behavior at a fundamental
level. Nevertheless, in spite of a quarter of a century of intensive
research into the subject, some aspects of their behavior still remain
puzzling.

When a QPC is opened up by sweeping the gate voltage, $V_\mathrm{g}$,
that controls its width, its linear conductance famously rises in
integer steps of the conductance quantum, $\GQ = 2 e^2/h$
\cite{Wharam1988,Wees1988}. This conductance quantization is well
understood \cite{Buettiker1990} and constitutes one of
the foundations of mesoscopic physics. However, during the first
conductance step, where the dimensionless conductance $g = G/\GQ$
changes from 0 to 1 (``closed'' to ``open'' QPC), an unexpected
shoulder is generically observed near $g \simeq 0.7$.  More
generally, the conductance shows anomalous behavior as function of
temperature ($T$), magnetic field ($B$) and source-drain voltage
($\Vsd$) throughout the regime $0.5 \lesssim g \lesssim 0.9$, where
the QPC is ``subopen''. The source of this behavior, collectively
known as the ``0.7-anomaly'', has been controversially discussed
\cite{Thomas1996,*Thomas1998,Reilly2002,Reilly2005,Jaksch2006,
  Koop2007,Smith2011,Iqbal2013,Brun2014,Cronenwett2002,Meir2002,*Hirose2003,*Golub2006,*Rejec2006,Chen2008,Chen2010a,Chen2012,Potok2002,Cronenwett2002,Meir2002,*Hirose2003,*Golub2006,*Rejec2006,Iqbal2013,Wang1996,*Wang1998,Wang1996,*Wang1998,Brun2014,Micolich2011,Komijani2009,Chung2007,Bauer2013,*Heyder2016} ever since it
was first systematically described in 1996 \cite{Thomas1996}.  Though
no consensus has yet been reached regarding its detailed microscopic
origin \cite{Iqbal2013,Bauer2013}, general agreement exists that it
involves electron spin dynamics and geometrically-enhanced interaction
effects. 

In this paper we further explore the van Hove ridge scenario, proposed
in \cite{Bauer2013}. It asserts that the $0.7$ anomaly is a direct
consequence of a ``van Hove ridge'', i.\,e.\ a smeared van Hove peak
in the energy-resolved local density of states (LDOS)
$\mathcal A_i(\omega)$ at the bottom of the lowest $1$D subband of the
QPC. Its shape follows that of the QPC barrier
\cite{Bauer2013,Bauer2014,*Heyder2015,*Heyder2016} and in the subopen regime, where
the barrier top lies just below the chemical potential $\mu$, it
causes the LDOS \textit{at $\mu$} to be strongly enhanced. This
reflects the fact that electrons slow down while crossing
the QPC barrier (since the semiclassical velocity of an electron with
energy $\omega$ at position $i$ is inversely proportional to the
LDOS, $\mathcal A_i (\omega) \sim v^{-1}$).  The slow electrons experience
strongly enhanced mutual interactions, with striking consequences for
various physical properties.

In this paper, we elucidate their effect on various
\textit{dynamical} quantities, which we extract from real-frequency
correlation functions computed using the functional Renormalization
Group (fRG) on the Keldysh contour   
\cite{Jakobs2010,Jakobs2010-2,Karrasch2006a,Metzner2012}.  We
compute (i) the frequency dependence of the LDOS, finding that its
maximum is pinned to $\mu$ in the subopen regime, indicative of a
Coulomb-blockade type behaviour; (ii) the temperature dependence of
the linear conductance, finding qualitative agreement with experiment;
(iii) the dynamical spin susceptibility $\chi (\omega)$, from which we
extract a characteristic time scale $t_{\rm spin}$ for spin
fluctuations, and (iv) the time $t_{\rm trav}$ for a quasiparticle to
traverse the QPC, which we extract from the single-particle scattering
matrix $S(\omega)$.  Intermediate interaction strengths suffice to
obtain the characteristic 0.7 shoulder at finite temperatures. We find
strong links between the $\omega$-dependence of the spin
susceptibility, the one-particle S-matrix, and the form of the LDOS.
As long as the van Hove ridge is pinned to $\mu$, interactions cause
relevant degrees of freedom to slow down, inducing significant
increases in both $t_{\rm trav}$ and $t_{\rm spin}$. Moreover, these
two times are comparable in magnitude, implying that a quasiparticle
traversing the QPC encounters a quasi-static spin background. 
 This provides a link to other proposed explanations
of the $0.7$ anomaly
\cite{Thomas1996,*Thomas1998,Reilly2002,Koop2007,Smith2011,Wang1996,*Wang1998,Reilly2005,Jaksch2006,Chen2008,Chen2010a,Chen2012,Potok2002,Cronenwett2002,Meir2002,*Hirose2003,*Golub2006,*Rejec2006,Iqbal2013,Brun2014}.

\textit{Model.}---We model the QPC by a smooth potential barrier
describing the effective 1D-potential along the transport
direction. Information about the channel's transverse structure 
is incorporated into space-dependent model parameters. After
discretizing the longitudinal position coordinate as $x=ai$, with site
index $i$ and lattice spacing $a$, the model Hamiltonian has the form
\cite{Bauer2013}
\begin{equation}
\label{eq:define_H}
	\mathcal H = - \sum_{\sigma, i} \tau_{i} \left( c^\dagger_{i+1,\sigma} c_{i,\sigma} + {\rm h.c.} \right) + \sum_{i} U_i c^\dagger_{i\uparrow} c_{i\uparrow} c^\dagger_{i\downarrow} c_{i\downarrow} .
\end{equation}
It describes an infinite tight-binding chain with nearest-
neighbor hopping $\tau_i$ of quasiparticles with spin $\sigma = \uparrow, \downarrow$ and short-range interactions $U_i$.
The hopping amplitude $\tau_i$
varies smoothly with $i$, thus creating an effective potential barrier
$V_i = -(\tau_i+\tau_{i+1})+2 \tau$
measured w.r.t. the leads' band bottom $-2\tau$.
We choose $U_i \neq 0$ and $\tau_i \neq \tau$
only for $N = 2N'+1$ sites, symmetric around $i=0$,
that define the extent of the QPC (central region).
$U_i$ is constant in the center of the QPC with
$U_0 = U$ and drops smoothly to zero 
as $i$ approaches the edges of the central region at sites $\pm N'$.
We tune the hopping such that the effective barrier is symmetric and parabolic near the top, 
$V_i = \tilde \Vg - i^2 \wx^2/(4\tau)$, where the barrier height $\tilde \Vg$ mimics the role of gate voltage from
experiment, and the curvature $\wx$ sets the characteristic length scale $\lx = a \sqrt{\tau/ \wx}$ of the QPC.
We vary $\tilde \Vg$ such that the barrier crosses the chemical potential $\mu$.
The precise form of $U_i$ and $\tau_i$ is given in \cite{supplement}.
The model is solved with the perturbatively-truncated Keldysh-fRG in equilibrium \cite{supplement}.
The plots shown are computed for $\tau= 1$, $U=0.7 \tau$, $\mu = -1.475 \tau$, $\Vg = \tilde \Vg - \mu -2 \tau \in [-2.83, 1.83] \wx$, and $\wx \approx 0.03 \tau$ [with $\hbar = 1$]. 

\textit{Local density of states.}---\begin{figure}
\includegraphics[width=\columnwidth]{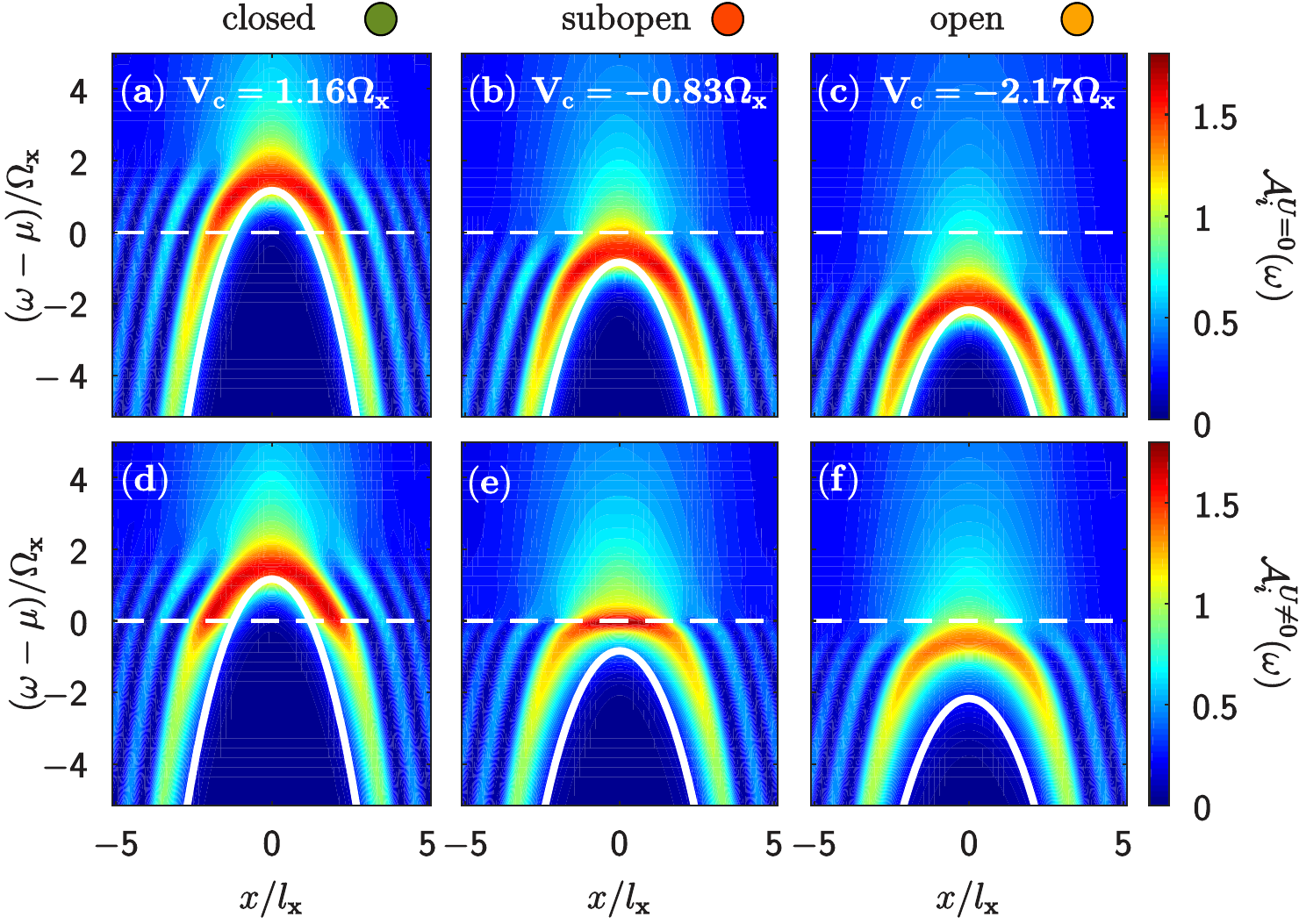}
\vspace{-5mm}

\caption{van Hove ridge in the LDOS ${\cal A}_i(\omega)$ (color scale) of a non-interacting (upper row) and interacting (lower row) QPC, plotted as function of energy $\omega-\mu$ and position $x = ai$. The thick solid white line depicts the effective bare potential barrier $V_i$, the thin dashed white line the chemical potential $\mu$. From left to right: closed, subopen and open regimes. With interactions, the Hove ridge is shifted upward and flattened in the (sub-)open regime [compare (b) and (e), (c) and (f)].
\vspace{-5mm}}
\label{fig:LDOS}
\end{figure}
\begin{figure}
\includegraphics[width=\columnwidth]{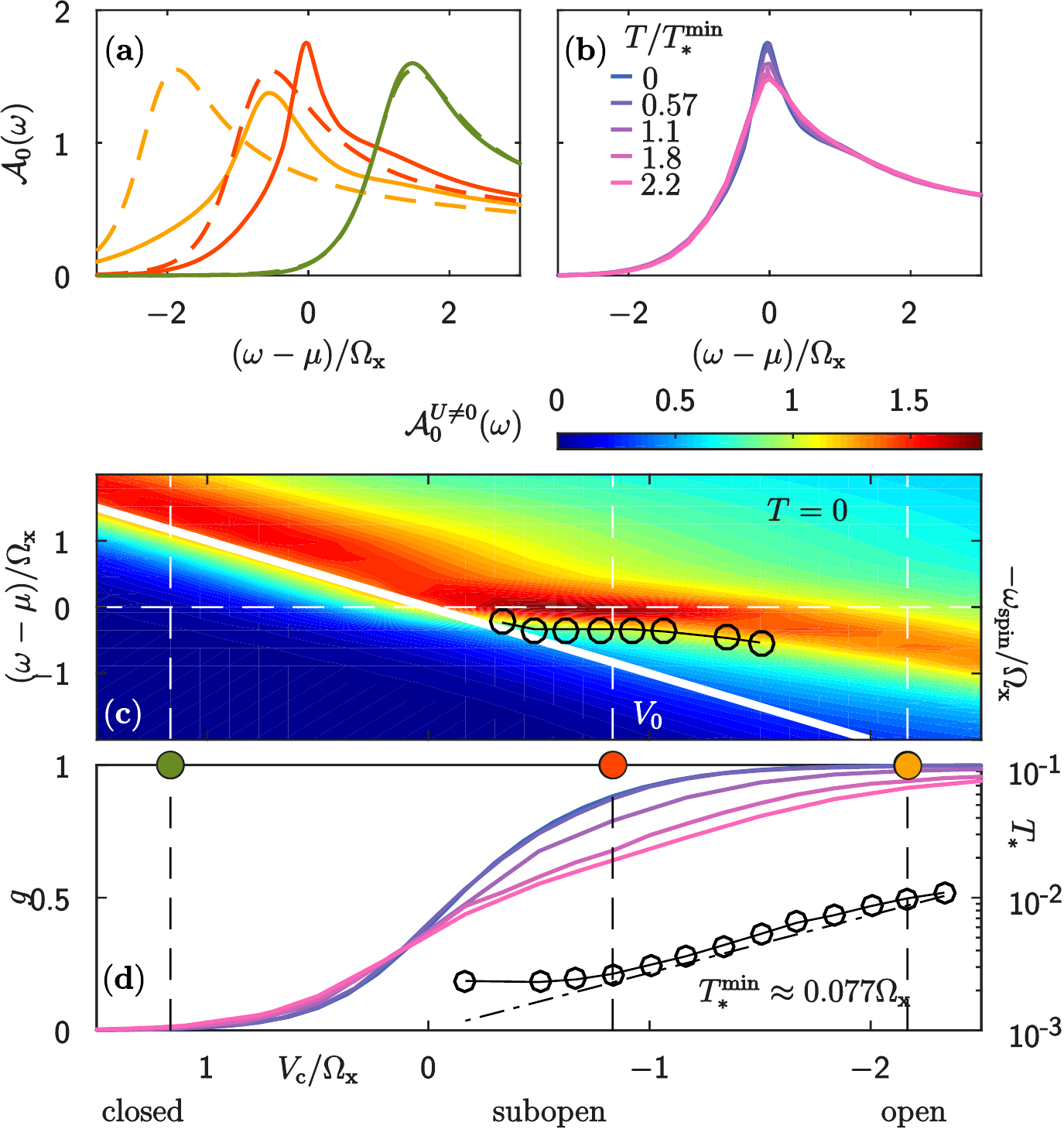}
\caption{ (a) The interacting LDOS (solid lines) and bare LDOS (dashed lines), plotted as function
  of energy $\omega$ for three values of $\Vg$, indicated by dots of
  corresponding color in (c,d). In the subopen (red) and open (orange)
  regimes, interactions shift the van Hove peak to larger frequencies,
  as the barrier height is renormalized. Moreover, in the subopen
  regime, flattening of the van Hove ridge causes the peak to become
  sharper and higher.  (b) ${\cal A}_0 (\omega)$ in the subopen
  regime, for three different temperatures. At larger temperatures,
  the maximum is lower as weight is shifted into the flanks of the van
  Hove ridge and redistributed in the band.  (c)
  ${\cal A}_{0} (\omega)$, the interacting LDOS (color scale) at the
  central site, as function of $\omega$ and $\Vg$. The solid white
  line shows the bare barrier height, $V_0$. In the subopen regime the
  energy of the van Hove ridge maximum, $\omega_{\rm max}$, is pinned
  to the chemical potential. The black circles show the characteristic
  frequency $\omega_{\rm spin}$ of the spin susceptibility
  $\chi$. They clearly follow the LDOS maximum.  (d) Conductance $g$
  (left axis) for different temperatures, and $T_*$ (circles), as
  defined in Eq. \eqref{eq:define_Tstar}, on a logarithmic scale
  (right axis). Temperature is measured in units of
  $T_*^{\rm min} = {\rm min}~T_* (\Vg)$. As guide to the eye:
  $0.001 \cdot \exp(-\Vg/\wx)$ (dashed-dotted line). 
\vspace{-5mm}}
\label{fig:resonant_energy_structure}
\end{figure}
It has been argued in Ref. \cite{Bauer2013} that the physics of the QPC is governed by the LDOS,
$
	{\cal A}_i (\omega) = -\frac 1\pi {\rm Im} G^{R}_{ii} (\omega),
$
where $G^R_{ij}$ is the retarded single-particle Green's function between site $i$ and $j$.
Fig.~\ref{fig:LDOS}(a-c) shows the bare LDOS ${\cal A}^{U=0}_i(\omega)$
of the QPC as a function of site $i$ and frequency $\omega$ at three values of the barrier height $\Vg$. The bare LDOS has a
maximum just above the band bottom, visible as a red structure, that follows the shape of the effective potential (thick white
line). This structure is the bare van Hove
ridge discussed in \cite{Bauer2013}, the apex of the which has a maximum value $\sim \left( \wx \tau \right)^{-1/2}$, and occurs at an energy $\omega_{\rm max} (\Vg)$ that lies slightly higher
than the bare potential maximum $V_0$, by an amount $\sim \wx$.

Upon adding interactions, we obtain Fig.~\ref{fig:LDOS}(d-f), which shows two striking differences to the non-interacting case: In the (sub-)open regime the renormalized van Hove ridge is shifted upwards in energy ($\omega_{\rm max}$ is larger) and becomes flatter spatially. Both of these effects may \textit{qualitatively} be understood by a mean field argument \cite{PhysRevB.79.235313, PhysRevB.88.075305}: 
The slope of the van Hove ridge may be interpreted as reflecting the shape of an effective, renormalized potential barrier, which is shifted upwards relative to the bare barrier by a Hartree-shift proportional to the local electron density. 
Away from the center, the density is higher, such that the shift is larger, causing the van Hove ridge to become flatter as function of $x$ near its apex, while becoming narrower and higher as function of $\omega$. This is also seen clearly in Fig.~\ref{fig:resonant_energy_structure}(a), which shows the interacting (solid lines) and bare (dashed lines) LDOS ${\cal A}_0 (\omega)$. The $x$-flattening and $\omega$-sharpening is most striking in the subopen regime, where the van Hove ridge apex intersects the chemical potential [Fig.~\ref{fig:LDOS}(e)], because there the interaction-induced effects are largest.
We have checked our Keldysh-fRG results against DMRG computations of the system with somewhat different parameters \cite{supplement}, finding good qualitative agreement and, in particular, the same values for $\omega_{\rm max}$.

The evolution of ${\cal A}_0 (\omega)$ as $\Vg$ is varied is shown in
Fig.~\ref{fig:resonant_energy_structure}(c). As $\Vg$ is lowered, the
energy $\omega_{\rm max}$ of the Hove ridge maximum follows the bare
barrier top (solid white line) as long as the QPC is closed, then
remains \textit{pinned} at the chemical potential throughout the
subopen regime to form a plateau-like structure, and finally decreases
again only deep in the open regime (compare Fig.~1(d) of
\cite{PhysRevB.79.235313}).  We interpret this plateau-like
structure as a precursor of Coulomb blockade behavior, since it
arises from the interactions of electrons in a region of limited
spatial extent.  
   
\textit{Finite temperature.}---
This structure sheds new light on the temperature dependence of
the linear conductance on temperature. When the temperature, $T$, is
increased, the van Hove peak in the LDOS retains its overall shape
and is broadened only slightly (for $T \lesssim \wx/10$)
[Fig.~\ref{fig:resonant_energy_structure}(b)].  At the same time,
the first conductance step is flattened out in a characteristic,
asymmetric fashion [Fig.~\ref{fig:resonant_energy_structure}(d)], in
qualitative agreement with experiment (compare Fig.~2(f) of
Ref.~\cite{Bauer2013}).  This can be understood as follows
\cite{Bauer2013}: Increasing $T$ increases the available phase
space for inelastic scattering, thus enhancing interaction 
effects. Their strength is governed by the LDOS near the chemical
potential, which is particularly large \textit{throughout the subopen
  region}, due to the pinning of $\omega_{\max}$ to the chemical  
potential. Accordingly, interaction-induced backscattering is large in
the whole subopen regime, leading to a strong suppression of the       
conductance [Fig.~\ref{fig:resonant_energy_structure}(d)] even into
the open regime.  At pinch-off, the conductance is slightly increased
due to thermal activation.

To quantify the strength of the temperature dependence as function of $\Vg$, we expand
the conductance as
\begin{equation}
\label{eq:define_Tstar}
	g (T, \Vg) = g (0, \Vg) - \frac{T^2}{T_*^2(\Vg)} + {\cal O}(T^3),
\end{equation}
as appropriate for a Fermi liquid \cite{Bauer2013}.
The $T_*(\Vg)$ values extracted from our finite-$T$ data [see Fig.~\ref{fig:resonant_energy_structure}(d), circles] depend roughly exponentially on gate voltage $T_*(\Vg) \sim \exp(-\Vg/\wx)$ [Fig.~\ref{fig:resonant_energy_structure}(d), dashed-dotted line], when the QPC is tuned from subopen to open, reflecting the $\Vg$-dependence of the bare QPC transmission rate \cite{Bauer2013}.

\textit{Spin susceptibility.}---
\begin{figure}
\includegraphics[width=\columnwidth]{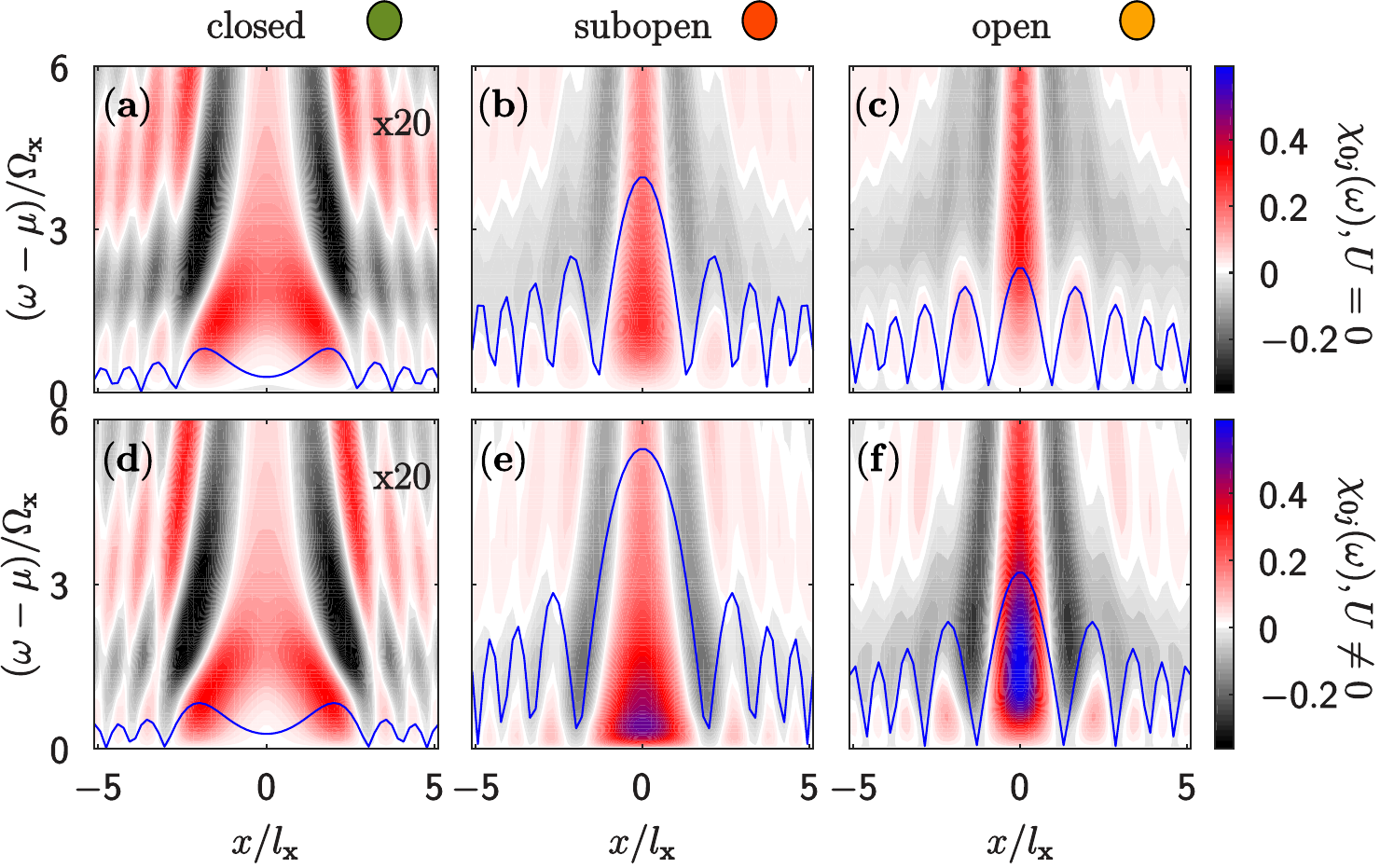}\vspace{-3mm}

\caption{Non-interacting (a-c) and interacting (d-f) dynamical spin susceptibility [multiplied by a factor of $20$ in order to be visible in (a) and (d)], for a closed, subopen and open QPC. The blue line shows $\vert {\rm Im} \left(G_{0i} (\omega=\mu) \right) \vert$ (a.u.).}
\label{fig:spin_spin_interacting}
\end{figure}
\begin{figure}
\includegraphics[width=\columnwidth]{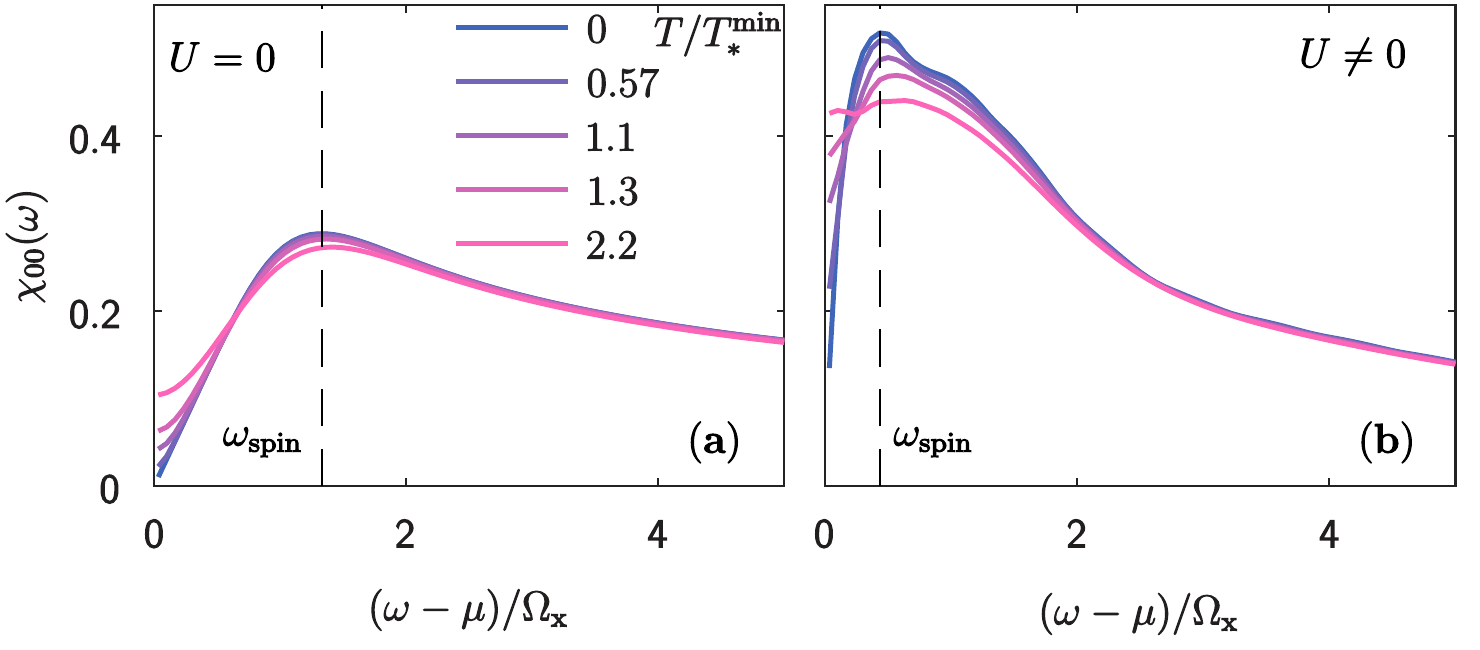}\vspace{-3mm}

\caption{Non-interacting (a) and interacting (b) spin-spin correlations on the central site in the subopen regime at different temperatures, i.e. the blue lines are vertical cuts of Fig.~\ref{fig:spin_spin_interacting}(b), (e) through $x=0$. The dashed black line is at $\omega=\omega_{\rm spin}$. The shoulder in (b) is due to the LDOS-dependent enhancement of the spin susceptibility due to interactions. \vspace{-3mm}}
\label{fig:spin_spin_cuts}
\end{figure}
In the van Hove ridge scenario a key property of a subopen QPC is the presence of ``slow spin fluctuations'' \cite{Bauer2013}, as advocated also
in Ref.~\cite{Aryanpour2009}.    
 To explore this, we have computed the dynamical equilibrium spin susceptibility
\begin{equation}  
\label{eq:spin-noise_definition}
	\chi_{ij} (\omega) = \int dt \langle {\cal T} S^z_i (t) S^z_j (0) \rangle \exp(i \omega t),
\end{equation}
where $\cal T$ denotes time-ordering.
In a Fermi liquid, the spin susceptibility is determined by the particle-hole bubble and thus governed by single-particle properties.
However, due to the inhomogenuity of the QPC, both the frequency- and position-dependence of the spin susceptibility are non-trivial.
For now, we focus on $\chi_{0j}$, shown in Fig.~\ref{fig:spin_spin_interacting}, which has the following salient features:

(i) $\chi_{0j}$ oscillates with a spatially varying wavelength, which becomes shorter as the QPC is opened or the energy increased. 
For small frequencies $\omega$ the wavelength of these oscillations is determined by the ``local Fermi wavelength'' $\lambda_F$, which can be extracted from $\vert {\rm Im} \, G^R_{0j} (\mu) \vert$ (blue line in Fig.~\ref{fig:spin_spin_interacting}). 
In the subopen regime, $\lambda_{\rm F}$ is large in the center, where the density is small, such that the sign of the spin susceptibility only changes far away from the center. Thus, an excited spin in the center leads to a rather large cloud (covering a region of $\sim 3 \lx$) of co-oriented spins.
Away from the QPC the oscillations in $\chi_{0j}$ simply follow the Friedel oscillations.

(ii) On the central site, $\chi_{00}(\omega)$ shows a clear characteristic at a frequency $\omega_{\rm spin}(\Vg)$, whose dependence on $\Vg$ follows that of $\omega_{\rm max}$ [$-\omega_{\rm spin}$ is indicated by black circles in Fig.~\ref{fig:resonant_energy_structure}(c)].
In general, for small energies, $\omega_{\rm spin}$ is set by the distance between the chemical potential and the nearest peak in the LDOS \cite{supplement}.

(iii) The spin susceptibility $\chi_{0i}(\omega)$ is amplified by interactions (Stoner physics) [compare Fig.~\ref{fig:spin_spin_interacting}(a-c) and Fig.~\ref{fig:spin_spin_interacting}(d-f); also Fig.~\ref{fig:spin_spin_cuts}(a) and (b)]. 
Interactions also amplify the temperature-induced reduction of the spin susceptibility at $\omega_{\rm spin}$ [Fig.~\ref{fig:spin_spin_cuts}(a,b)]. This effect is of similar strength as the decrease of the LDOS at $\omega_{\rm max}$ [Fig.~\ref{fig:resonant_energy_structure}(b)].

\textit{Traversal time.}---
\begin{figure}
\includegraphics[width=\columnwidth]{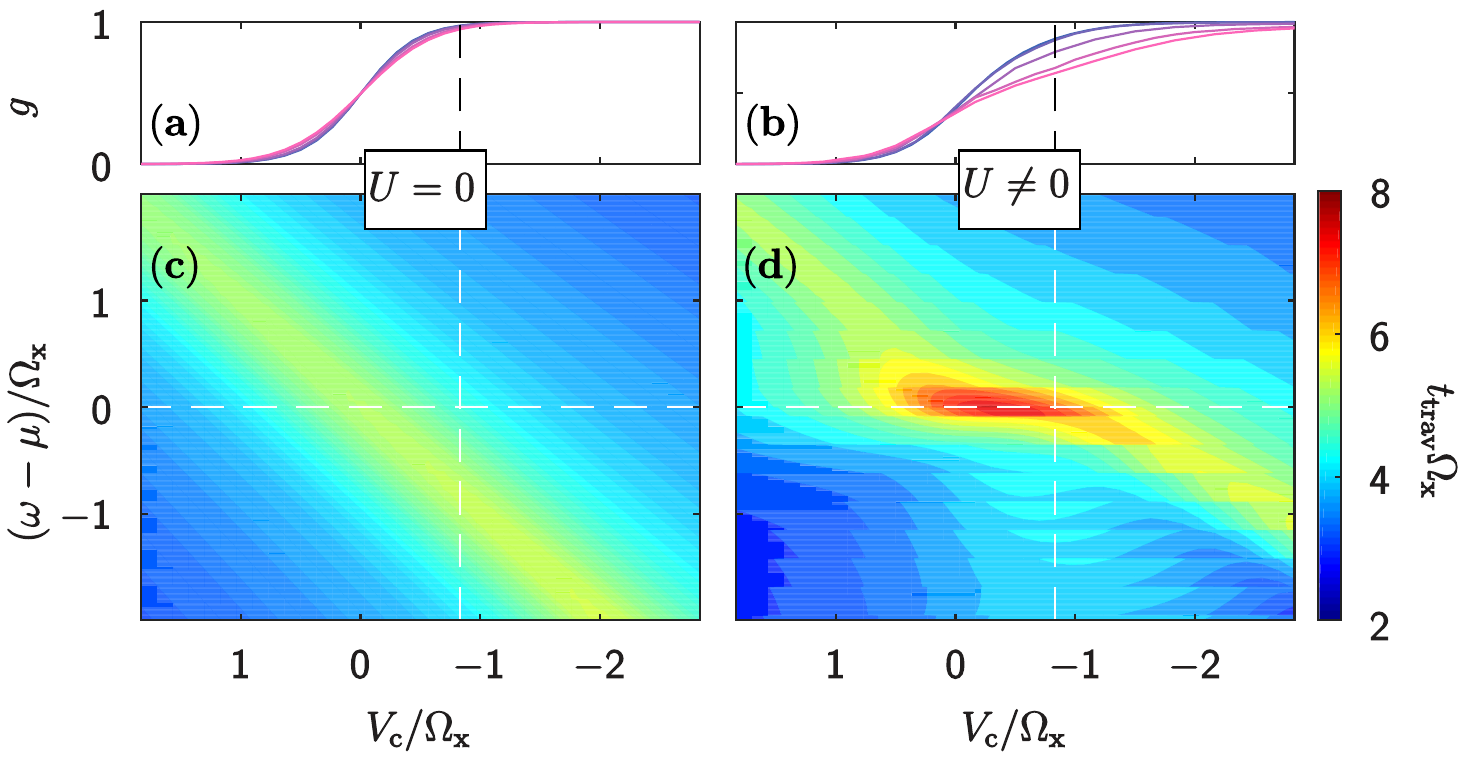}\vspace{-3mm}

\caption{Comparison of non-interacting (a,c) and interacting (b,d)
  traversal time. (a,b): Conductance $g$ as function of gate voltage
  $\Vg$, to identificy closed, subopen and open regimes. The
  color code is identical to
  Fig.~\ref{fig:resonant_energy_structure}; (c,d): Traversal time
  [Eq. \eqref{eq:time_traversal}] as function of frequency $\omega$
  and gate voltage $\Vg$. 
  While the traversal time of modes below the barrier is small, these modes have low transmission probability and are irrelevant when determining the timescale of transport.
  \vspace{-3mm}}
\label{fig:traversal_time}
\end{figure}
The traversal time $t_{\rm trav}$ for a single incident quasiparticle with energy $\omega$ to traverse a scattering region can be obtained by a procedure due to Wigner \cite{Wigner1955}, which relates it to the scattering-induced dispersion of the incident wave-packet: It is given by
\begin{equation}
\label{eq:time_traversal}
	t_{\rm trav} (\omega) = t_0 (\omega) + t_{\rm delay}(\omega),\, t_{\rm delay}(\omega) = 2 \partial_\omega \phi (\omega),
\end{equation}
where $t_0 (\omega)$ is the traversal time through the central region
with the potential and interactions being turned off, $t_{\rm delay}$
and $\phi(\omega)$ are the delay time and the scattering phase shift
due to the potential- and interaction-induced slow-down of the
quasiparticles.  In our setup $\phi(\omega)$ is the phase of the
left-right-component of the zero-temperature single-particle S-matrix,
\begin{equation}
\label{eq:S-Matrix}
	S_{\rm l,r}(\omega) = - 2\pi i \tau \rho(\omega) G^R_{-N', N'} (\omega),
\end{equation}
where $\rho(\omega)$ is the lead density of states at the sites $\pm (N'+1)$ in the absence of the central region and $\tau$ is the hopping amplitude there. $\vert S_{\rm l,r} (\omega) \vert^2$ yields the transmission probability.
Figs.~\ref{fig:traversal_time}(a,b) show the traversal time. Though calculated from a non-local correlation function, its behaviour is strikingly similar to that of the LDOS at the central site, Fig.~\ref{fig:resonant_energy_structure}(c). This is consistent with the semiclassical interpretation ${\cal A} \sim v^{-1}$: Whenever the LDOS is large, quasiparticles are slow and thus a large time is required to traverse the QPC. 

Interestingly, we find that in the subopen regime the traversal time $t_{\rm trav}$ is of the same order as the characteristic time scale, $t_{\rm spin} = \frac{2 \pi}{\omega_{\rm spin}}$, associated with spin fluctuations, namely  $t_{\rm trav} \lesssim 8/\wx$ and $t_{\rm spin} \lesssim 10 /\wx$. We note that with our parameters, $t_{0} \approx 1.3 /\wx$, thus $t_{\rm trav}$ is dominated by the delay time.
That $t_{\rm trav}$ and $t_{\rm spin}$ are comparable in magnitude is
consistent with a Fermi-liquid description of the system (which
underlies the fRG-method used here): The only stable degrees of
freedom in a Fermi liquid are dressed electron- and hole-like
quasiparticles, and spin fluctuations arise via electron-hole-like
excitations. Near the QPC center ($x \lesssim \lx$) the lifetime of
spin fluctuations is thus governed by the quasiparticle decay
time. Heuristically, this roughly corresponds to $t_{\rm trav}$, as
the region where interaction effects are strongest extends over only
few $\lambda_F$-oscillations. Though we find no static
contributions to the dynamical spin susceptibility at zero magnetic
field, the fact that $t_{\rm spin} \simeq t_{\rm trav}$, together with
the extended spatial structure of the spin susceptibility in the subopen
regime, suggests the heuristic view that a quasiparticle traversing
the QPC encounters a quasi-static, spatially coherent spin environment.

\textit{Conclusions.}---Our results allow us to establish contact
with two other prominent scenarios that have been proposed to explain
the $0.7$ anomaly. (i) According to the ``spin-polarization
scenario'', interactions cause the spin degree of freedom in the QPC
to spontaneously polarize, giving rise to a non-zero magnetization
even at vanishing magnetic field, $B=0$ 
\cite{Thomas1996,*Thomas1998,Reilly2002,Koop2007,Smith2011,Wang1996,*Wang1998,Reilly2005,Jaksch2006,Chen2008,Chen2010a,Chen2012,Potok2002}.
(ii) According to the ``quasi-localized spin scenario'' proposed by
Meir and coworkers \cite{Meir2002,*Hirose2003,*Golub2006,*Rejec2006}, a
subopen QPC hosts a quasi-localized state involving a spin-$\frac 12$
magnetic moment, causing Kondo-like conductance anomalies
\cite{Cronenwett2002,Meir2002,*Hirose2003,*Golub2006,*Rejec2006,Iqbal2013,Brun2014}. At      
low energies, a quasi-localized spin would be screened, giving rise to
Fermi-liquid behavior that includes slow spin fluctuations.
These two scenarios thus seem to offer starkly contrasting views of
the spin structure in a QPC: (i) spatially extended but static in
time, vs.\ (ii) spatially localized but fluctuating in time. Our work
suggests that a view that entails elements of both: the spin structure
fluctuates in time, in accord with (ii), but \textit{slowly} -- which
is compatible with (i) if one is willing to reinterpret ``spontaneous
polarization'' as ``slowly fluctuating polarization''. And the spin
structure is spatially coherent, in accord with (i), over a region of
\textit{finite extent} -- which is compatible with (ii) if one is
willing to associate a nonzero spatial extent and a finite life-time
with the quasi-localized state evoked there. We thus suggest that the
controversy between the opposing views (i) and (ii) can be resolved by
associating the quasi-localized state evoked in (ii) with the slow
electrons of the van Hove ridge, and realizing that these constitute a
quasi-static, spatially coherent spin environment, in the spirit of
(i), for electrons traversing the QPC. Thus, though the various
scenarios differ substantially in their details (and if one insists on
comparing these the controversy will never be put to rest), they can
be argued to have a common core: a \textit{slowly fluctuating spin
  structure of finite spatial extent} in the center of the
QPC. Moreover, our work, shows that this spin structure originates
naturally from the same interplay of interactions and QPC barrier
geometry, encoded in the van Hove ridge, that causes transport
properties to be anomalous.
   
We thank F. Bauer, J. Heyder, Y. Meir and L. Weidinger for useful discussions. BB thanks S. R. White for discussions on the DMRG setup. The authors are supported by the DFG through the Excellence Cluster ``Nanosystems Initiative Munich'', SFB/TR 12, SFB 631. 

\vspace{-5mm}

\bibliography{bibliography}

\begin{thebibliography}{51}%
\makeatletter
\providecommand \@ifxundefined [1]{%
 \@ifx{#1\undefined}
}%
\providecommand \@ifnum [1]{%
 \ifnum #1\expandafter \@firstoftwo
 \else \expandafter \@secondoftwo
 \fi
}%
\providecommand \@ifx [1]{%
 \ifx #1\expandafter \@firstoftwo
 \else \expandafter \@secondoftwo
 \fi
}%
\providecommand \natexlab [1]{#1}%
\providecommand \enquote  [1]{``#1''}%
\providecommand \bibnamefont  [1]{#1}%
\providecommand \bibfnamefont [1]{#1}%
\providecommand \citenamefont [1]{#1}%
\providecommand \href@noop [0]{\@secondoftwo}%
\providecommand \href [0]{\begingroup \@sanitize@url \@href}%
\providecommand \@href[1]{\@@startlink{#1}\@@href}%
\providecommand \@@href[1]{\endgroup#1\@@endlink}%
\providecommand \@sanitize@url [0]{\catcode `\\12\catcode `\$12\catcode
  `\&12\catcode `\#12\catcode `\^12\catcode `\_12\catcode `\%12\relax}%
\providecommand \@@startlink[1]{}%
\providecommand \@@endlink[0]{}%
\providecommand \url  [0]{\begingroup\@sanitize@url \@url }%
\providecommand \@url [1]{\endgroup\@href {#1}{\urlprefix }}%
\providecommand \urlprefix  [0]{URL }%
\providecommand \Eprint [0]{\href }%
\providecommand \doibase [0]{http://dx.doi.org/}%
\providecommand \selectlanguage [0]{\@gobble}%
\providecommand \bibinfo  [0]{\@secondoftwo}%
\providecommand \bibfield  [0]{\@secondoftwo}%
\providecommand \translation [1]{[#1]}%
\providecommand \BibitemOpen [0]{}%
\providecommand \bibitemStop [0]{}%
\providecommand \bibitemNoStop [0]{.\EOS\space}%
\providecommand \EOS [0]{\spacefactor3000\relax}%
\providecommand \BibitemShut  [1]{\csname bibitem#1\endcsname}%
\let\auto@bib@innerbib\@empty
\bibitem [{\citenamefont {Wharam}\ \emph {et~al.}(1988)\citenamefont {Wharam},
  \citenamefont {Thornton}, \citenamefont {Newbury}, \citenamefont {Pepper},
  \citenamefont {Ahmed}, \citenamefont {Frost}, \citenamefont {Hasko},
  \citenamefont {Peacock}, \citenamefont {Ritchie},\ and\ \citenamefont
  {Jones}}]{Wharam1988}%
  \BibitemOpen
  \bibfield  {author} {\bibinfo {author} {\bibfnamefont {D.~A.}\ \bibnamefont
  {Wharam}}, \bibinfo {author} {\bibfnamefont {T.~J.}\ \bibnamefont
  {Thornton}}, \bibinfo {author} {\bibfnamefont {R.}~\bibnamefont {Newbury}},
  \bibinfo {author} {\bibfnamefont {M.}~\bibnamefont {Pepper}}, \bibinfo
  {author} {\bibfnamefont {H.}~\bibnamefont {Ahmed}}, \bibinfo {author}
  {\bibfnamefont {J.~E.~F.}\ \bibnamefont {Frost}}, \bibinfo {author}
  {\bibfnamefont {D.~G.}\ \bibnamefont {Hasko}}, \bibinfo {author}
  {\bibfnamefont {D.~C.}\ \bibnamefont {Peacock}}, \bibinfo {author}
  {\bibfnamefont {D.~A.}\ \bibnamefont {Ritchie}}, \ and\ \bibinfo {author}
  {\bibfnamefont {G.~A.~C.}\ \bibnamefont {Jones}},\ }\href
  {http://stacks.iop.org/0022-3719/21/i=8/a=002} {\bibfield  {journal}
  {\bibinfo  {journal} {Journal of Physics C: Solid State Physics}\ }\textbf
  {\bibinfo {volume} {21}},\ \bibinfo {pages} {L209} (\bibinfo {year}
  {1988})}\BibitemShut {NoStop}%
\bibitem [{\citenamefont {van Wees}\ \emph {et~al.}(1988)\citenamefont {van
  Wees}, \citenamefont {van Houten}, \citenamefont {Beenakker}, \citenamefont
  {Williamson}, \citenamefont {Kouwenhoven}, \citenamefont {van~der Marel},\
  and\ \citenamefont {Foxon}}]{Wees1988}%
  \BibitemOpen
  \bibfield  {author} {\bibinfo {author} {\bibfnamefont {B.~J.}\ \bibnamefont
  {van Wees}}, \bibinfo {author} {\bibfnamefont {H.}~\bibnamefont {van
  Houten}}, \bibinfo {author} {\bibfnamefont {C.~W.~J.}\ \bibnamefont
  {Beenakker}}, \bibinfo {author} {\bibfnamefont {J.~G.}\ \bibnamefont
  {Williamson}}, \bibinfo {author} {\bibfnamefont {L.~P.}\ \bibnamefont
  {Kouwenhoven}}, \bibinfo {author} {\bibfnamefont {D.}~\bibnamefont {van~der
  Marel}}, \ and\ \bibinfo {author} {\bibfnamefont {C.~T.}\ \bibnamefont
  {Foxon}},\ }\href {\doibase 10.1103/PhysRevLett.60.848} {\bibfield  {journal}
  {\bibinfo  {journal} {Phys. Rev. Lett.}\ }\textbf {\bibinfo {volume} {60}},\
  \bibinfo {pages} {848} (\bibinfo {year} {1988})}\BibitemShut {NoStop}%
\bibitem [{\citenamefont {B\"uttiker}(1990)}]{Buettiker1990}%
  \BibitemOpen
  \bibfield  {author} {\bibinfo {author} {\bibfnamefont {M.}~\bibnamefont
  {B\"uttiker}},\ }\href {\doibase 10.1103/PhysRevB.41.7906} {\bibfield
  {journal} {\bibinfo  {journal} {Phys. Rev. B}\ }\textbf {\bibinfo {volume}
  {41}},\ \bibinfo {pages} {7906} (\bibinfo {year} {1990})}\BibitemShut
  {NoStop}%
\bibitem [{\citenamefont {Thomas}\ \emph {et~al.}(1996)\citenamefont {Thomas},
  \citenamefont {Nicholls}, \citenamefont {Simmons}, \citenamefont {Pepper},
  \citenamefont {Mace},\ and\ \citenamefont {Ritchie}}]{Thomas1996}%
  \BibitemOpen
  \bibfield  {author} {\bibinfo {author} {\bibfnamefont {K.~J.}\ \bibnamefont
  {Thomas}}, \bibinfo {author} {\bibfnamefont {J.~T.}\ \bibnamefont
  {Nicholls}}, \bibinfo {author} {\bibfnamefont {M.~Y.}\ \bibnamefont
  {Simmons}}, \bibinfo {author} {\bibfnamefont {M.}~\bibnamefont {Pepper}},
  \bibinfo {author} {\bibfnamefont {D.~R.}\ \bibnamefont {Mace}}, \ and\
  \bibinfo {author} {\bibfnamefont {D.~A.}\ \bibnamefont {Ritchie}},\ }\href
  {\doibase 10.1103/PhysRevLett.77.135} {\bibfield  {journal} {\bibinfo
  {journal} {Phys. Rev. Lett.}\ }\textbf {\bibinfo {volume} {77}},\ \bibinfo
  {pages} {135} (\bibinfo {year} {1996})}\BibitemShut {NoStop}%
\bibitem [{\citenamefont {Thomas}\ \emph {et~al.}(1998)\citenamefont {Thomas},
  \citenamefont {Nicholls}, \citenamefont {Appleyard}, \citenamefont {Simmons},
  \citenamefont {Pepper}, \citenamefont {Mace}, \citenamefont {Tribe},\ and\
  \citenamefont {Ritchie}}]{Thomas1998}%
  \BibitemOpen
  \bibfield  {author} {\bibinfo {author} {\bibfnamefont {K.~J.}\ \bibnamefont
  {Thomas}}, \bibinfo {author} {\bibfnamefont {J.~T.}\ \bibnamefont
  {Nicholls}}, \bibinfo {author} {\bibfnamefont {N.~J.}\ \bibnamefont
  {Appleyard}}, \bibinfo {author} {\bibfnamefont {M.~Y.}\ \bibnamefont
  {Simmons}}, \bibinfo {author} {\bibfnamefont {M.}~\bibnamefont {Pepper}},
  \bibinfo {author} {\bibfnamefont {D.~R.}\ \bibnamefont {Mace}}, \bibinfo
  {author} {\bibfnamefont {W.~R.}\ \bibnamefont {Tribe}}, \ and\ \bibinfo
  {author} {\bibfnamefont {D.~A.}\ \bibnamefont {Ritchie}},\ }\href {\doibase
  10.1103/PhysRevB.58.4846} {\bibfield  {journal} {\bibinfo  {journal} {Phys.
  Rev. B}\ }\textbf {\bibinfo {volume} {58}},\ \bibinfo {pages} {4846}
  (\bibinfo {year} {1998})}\BibitemShut {NoStop}%
\bibitem [{\citenamefont {Reilly}\ \emph {et~al.}(2002)\citenamefont {Reilly},
  \citenamefont {Buehler}, \citenamefont {O\char39{}Brien}, \citenamefont
  {Hamilton}, \citenamefont {Dzurak}, \citenamefont {Clark}, \citenamefont
  {Kane}, \citenamefont {Pfeiffer},\ and\ \citenamefont {West}}]{Reilly2002}%
  \BibitemOpen
  \bibfield  {author} {\bibinfo {author} {\bibfnamefont {D.~J.}\ \bibnamefont
  {Reilly}}, \bibinfo {author} {\bibfnamefont {T.~M.}\ \bibnamefont {Buehler}},
  \bibinfo {author} {\bibfnamefont {J.~L.}\ \bibnamefont {O\char39{}Brien}},
  \bibinfo {author} {\bibfnamefont {A.~R.}\ \bibnamefont {Hamilton}}, \bibinfo
  {author} {\bibfnamefont {A.~S.}\ \bibnamefont {Dzurak}}, \bibinfo {author}
  {\bibfnamefont {R.~G.}\ \bibnamefont {Clark}}, \bibinfo {author}
  {\bibfnamefont {B.~E.}\ \bibnamefont {Kane}}, \bibinfo {author}
  {\bibfnamefont {L.~N.}\ \bibnamefont {Pfeiffer}}, \ and\ \bibinfo {author}
  {\bibfnamefont {K.~W.}\ \bibnamefont {West}},\ }\href {\doibase
  10.1103/PhysRevLett.89.246801} {\bibfield  {journal} {\bibinfo  {journal}
  {Phys. Rev. Lett.}\ }\textbf {\bibinfo {volume} {89}},\ \bibinfo {pages}
  {246801} (\bibinfo {year} {2002})}\BibitemShut {NoStop}%
\bibitem [{\citenamefont {Reilly}(2005)}]{Reilly2005}%
  \BibitemOpen
  \bibfield  {author} {\bibinfo {author} {\bibfnamefont {D.~J.}\ \bibnamefont
  {Reilly}},\ }\href {\doibase 10.1103/PhysRevB.72.033309} {\bibfield
  {journal} {\bibinfo  {journal} {Phys. Rev. B}\ }\textbf {\bibinfo {volume}
  {72}},\ \bibinfo {pages} {033309} (\bibinfo {year} {2005})}\BibitemShut
  {NoStop}%
\bibitem [{\citenamefont {Jaksch}\ \emph {et~al.}(2006)\citenamefont {Jaksch},
  \citenamefont {Yakimenko},\ and\ \citenamefont {Berggren}}]{Jaksch2006}%
  \BibitemOpen
  \bibfield  {author} {\bibinfo {author} {\bibfnamefont {P.}~\bibnamefont
  {Jaksch}}, \bibinfo {author} {\bibfnamefont {I.}~\bibnamefont {Yakimenko}}, \
  and\ \bibinfo {author} {\bibfnamefont {K.-F.}\ \bibnamefont {Berggren}},\
  }\href {\doibase 10.1103/PhysRevB.74.235320} {\bibfield  {journal} {\bibinfo
  {journal} {Phys. Rev. B}\ }\textbf {\bibinfo {volume} {74}},\ \bibinfo
  {pages} {235320} (\bibinfo {year} {2006})}\BibitemShut {NoStop}%
\bibitem [{\citenamefont {Koop}\ \emph {et~al.}(2007)\citenamefont {Koop},
  \citenamefont {Lerescu}, \citenamefont {Liu}, \citenamefont {van Wees},
  \citenamefont {Reuter}, \citenamefont {A.D.Wieck},\ and\ \citenamefont
  {van~der Wal}}]{Koop2007}%
  \BibitemOpen
  \bibfield  {author} {\bibinfo {author} {\bibfnamefont {E.}~\bibnamefont
  {Koop}}, \bibinfo {author} {\bibfnamefont {A.}~\bibnamefont {Lerescu}},
  \bibinfo {author} {\bibfnamefont {J.}~\bibnamefont {Liu}}, \bibinfo {author}
  {\bibfnamefont {B.}~\bibnamefont {van Wees}}, \bibinfo {author}
  {\bibfnamefont {D.}~\bibnamefont {Reuter}}, \bibinfo {author} {\bibnamefont
  {A.D.Wieck}}, \ and\ \bibinfo {author} {\bibfnamefont {C.}~\bibnamefont
  {van~der Wal}},\ }\href {\doibase 10.1007/s10948-007-0289-5} {\bibfield
  {journal} {\bibinfo  {journal} {J. Supercond. Nov. Magn.}\ }\textbf {\bibinfo
  {volume} {20}},\ \bibinfo {pages} {433} (\bibinfo {year} {2007})}\BibitemShut
  {NoStop}%
\bibitem [{\citenamefont {Smith}\ \emph {et~al.}(2011)\citenamefont {Smith},
  \citenamefont {Hamilton}, \citenamefont {Thomas}, \citenamefont {Pepper},
  \citenamefont {Farrer}, \citenamefont {Griffiths}, \citenamefont {Jones},\
  and\ \citenamefont {Ritchie}}]{Smith2011}%
  \BibitemOpen
  \bibfield  {author} {\bibinfo {author} {\bibfnamefont {L.~W.}\ \bibnamefont
  {Smith}}, \bibinfo {author} {\bibfnamefont {A.~R.}\ \bibnamefont {Hamilton}},
  \bibinfo {author} {\bibfnamefont {K.~J.}\ \bibnamefont {Thomas}}, \bibinfo
  {author} {\bibfnamefont {M.}~\bibnamefont {Pepper}}, \bibinfo {author}
  {\bibfnamefont {I.}~\bibnamefont {Farrer}}, \bibinfo {author} {\bibfnamefont
  {J.~P.}\ \bibnamefont {Griffiths}}, \bibinfo {author} {\bibfnamefont
  {G.~A.~C.}\ \bibnamefont {Jones}}, \ and\ \bibinfo {author} {\bibfnamefont
  {D.~A.}\ \bibnamefont {Ritchie}},\ }\href {\doibase
  10.1103/PhysRevLett.107.126801} {\bibfield  {journal} {\bibinfo  {journal}
  {Phys. Rev. Lett.}\ }\textbf {\bibinfo {volume} {107}},\ \bibinfo {pages}
  {126801} (\bibinfo {year} {2011})}\BibitemShut {NoStop}%
\bibitem [{\citenamefont {Iqbal}\ \emph {et~al.}(2013)\citenamefont {Iqbal},
  \citenamefont {Levy}, \citenamefont {Koop}, \citenamefont {Dekker},
  \citenamefont {de~Jong}, \citenamefont {van~der Velde}, \citenamefont
  {Reuter}, \citenamefont {Wieck}, \citenamefont {Aguado}, \citenamefont
  {Meir},\ and\ \citenamefont {van~der Wal}}]{Iqbal2013}%
  \BibitemOpen
  \bibfield  {author} {\bibinfo {author} {\bibfnamefont {M.~J.}\ \bibnamefont
  {Iqbal}}, \bibinfo {author} {\bibfnamefont {R.}~\bibnamefont {Levy}},
  \bibinfo {author} {\bibfnamefont {E.~J.}\ \bibnamefont {Koop}}, \bibinfo
  {author} {\bibfnamefont {J.~B.}\ \bibnamefont {Dekker}}, \bibinfo {author}
  {\bibfnamefont {J.~P.}\ \bibnamefont {de~Jong}}, \bibinfo {author}
  {\bibfnamefont {J.~H.~M.}\ \bibnamefont {van~der Velde}}, \bibinfo {author}
  {\bibfnamefont {D.}~\bibnamefont {Reuter}}, \bibinfo {author} {\bibfnamefont
  {A.~D.}\ \bibnamefont {Wieck}}, \bibinfo {author} {\bibfnamefont
  {R.}~\bibnamefont {Aguado}}, \bibinfo {author} {\bibfnamefont
  {Y.}~\bibnamefont {Meir}}, \ and\ \bibinfo {author} {\bibfnamefont {C.~H.}\
  \bibnamefont {van~der Wal}},\ }\href {http://dx.doi.org/10.1038/nature12491}
  {\bibfield  {journal} {\bibinfo  {journal} {Nature}\ }\textbf {\bibinfo
  {volume} {501}},\ \bibinfo {pages} {79} (\bibinfo {year} {2013})}\BibitemShut
  {NoStop}%
\bibitem [{\citenamefont {Brun}\ \emph {et~al.}(2014)\citenamefont {Brun},
  \citenamefont {Martins}, \citenamefont {Faniel}, \citenamefont {Hackens},
  \citenamefont {Bachelier}, \citenamefont {Cavanna}, \citenamefont {Ulysse},
  \citenamefont {Ouerghi}, \citenamefont {Gennser}, \citenamefont {Mailly},
  \citenamefont {Huant}, \citenamefont {Bayot}, \citenamefont {Sanquer},\ and\
  \citenamefont {Sellier}}]{Brun2014}%
  \BibitemOpen
  \bibfield  {author} {\bibinfo {author} {\bibfnamefont {B.}~\bibnamefont
  {Brun}}, \bibinfo {author} {\bibfnamefont {F.}~\bibnamefont {Martins}},
  \bibinfo {author} {\bibfnamefont {S.}~\bibnamefont {Faniel}}, \bibinfo
  {author} {\bibfnamefont {B.}~\bibnamefont {Hackens}}, \bibinfo {author}
  {\bibfnamefont {G.}~\bibnamefont {Bachelier}}, \bibinfo {author}
  {\bibfnamefont {A.}~\bibnamefont {Cavanna}}, \bibinfo {author} {\bibfnamefont
  {C.}~\bibnamefont {Ulysse}}, \bibinfo {author} {\bibfnamefont
  {A.}~\bibnamefont {Ouerghi}}, \bibinfo {author} {\bibfnamefont
  {U.}~\bibnamefont {Gennser}}, \bibinfo {author} {\bibfnamefont
  {D.}~\bibnamefont {Mailly}}, \bibinfo {author} {\bibfnamefont
  {S.}~\bibnamefont {Huant}}, \bibinfo {author} {\bibfnamefont
  {V.}~\bibnamefont {Bayot}}, \bibinfo {author} {\bibfnamefont
  {M.}~\bibnamefont {Sanquer}}, \ and\ \bibinfo {author} {\bibfnamefont
  {H.}~\bibnamefont {Sellier}},\ }\href {\doibase doi:10.1038/ncomms5290}
  {\bibfield  {journal} {\bibinfo  {journal} {Nature Comm.}\ }\textbf {\bibinfo
  {volume} {5}},\ \bibinfo {pages} {4290} (\bibinfo {year} {2014})}\BibitemShut
  {NoStop}%
\bibitem [{\citenamefont {Cronenwett}\ \emph {et~al.}(2002)\citenamefont
  {Cronenwett}, \citenamefont {Lynch}, \citenamefont {Goldhaber-Gordon},
  \citenamefont {Kouwenhoven}, \citenamefont {Marcus}, \citenamefont {Hirose},
  \citenamefont {Wingreen},\ and\ \citenamefont {Umansky}}]{Cronenwett2002}%
  \BibitemOpen
  \bibfield  {author} {\bibinfo {author} {\bibfnamefont {S.~M.}\ \bibnamefont
  {Cronenwett}}, \bibinfo {author} {\bibfnamefont {H.~J.}\ \bibnamefont
  {Lynch}}, \bibinfo {author} {\bibfnamefont {D.}~\bibnamefont
  {Goldhaber-Gordon}}, \bibinfo {author} {\bibfnamefont {L.~P.}\ \bibnamefont
  {Kouwenhoven}}, \bibinfo {author} {\bibfnamefont {C.~M.}\ \bibnamefont
  {Marcus}}, \bibinfo {author} {\bibfnamefont {K.}~\bibnamefont {Hirose}},
  \bibinfo {author} {\bibfnamefont {N.~S.}\ \bibnamefont {Wingreen}}, \ and\
  \bibinfo {author} {\bibfnamefont {V.}~\bibnamefont {Umansky}},\ }\href
  {\doibase 10.1103/PhysRevLett.88.226805} {\bibfield  {journal} {\bibinfo
  {journal} {Phys. Rev. Lett.}\ }\textbf {\bibinfo {volume} {88}},\ \bibinfo
  {pages} {226805} (\bibinfo {year} {2002})}\BibitemShut {NoStop}%
\bibitem [{\citenamefont {Meir}\ \emph {et~al.}(2002)\citenamefont {Meir},
  \citenamefont {Hirose},\ and\ \citenamefont {Wingreen}}]{Meir2002}%
  \BibitemOpen
  \bibfield  {author} {\bibinfo {author} {\bibfnamefont {Y.}~\bibnamefont
  {Meir}}, \bibinfo {author} {\bibfnamefont {K.}~\bibnamefont {Hirose}}, \ and\
  \bibinfo {author} {\bibfnamefont {N.~S.}\ \bibnamefont {Wingreen}},\ }\href
  {\doibase 10.1103/PhysRevLett.89.196802} {\bibfield  {journal} {\bibinfo
  {journal} {Phys. Rev. Lett.}\ }\textbf {\bibinfo {volume} {89}},\ \bibinfo
  {pages} {196802} (\bibinfo {year} {2002})}\BibitemShut {NoStop}%
\bibitem [{\citenamefont {Hirose}\ \emph {et~al.}(2003)\citenamefont {Hirose},
  \citenamefont {Meir},\ and\ \citenamefont {Wingreen}}]{Hirose2003}%
  \BibitemOpen
  \bibfield  {author} {\bibinfo {author} {\bibfnamefont {K.}~\bibnamefont
  {Hirose}}, \bibinfo {author} {\bibfnamefont {Y.}~\bibnamefont {Meir}}, \ and\
  \bibinfo {author} {\bibfnamefont {N.~S.}\ \bibnamefont {Wingreen}},\ }\href
  {\doibase 10.1103/PhysRevLett.90.026804} {\bibfield  {journal} {\bibinfo
  {journal} {Phys. Rev. Lett.}\ }\textbf {\bibinfo {volume} {90}},\ \bibinfo
  {pages} {026804} (\bibinfo {year} {2003})}\BibitemShut {NoStop}%
\bibitem [{\citenamefont {Golub}\ \emph {et~al.}(2006)\citenamefont {Golub},
  \citenamefont {Aono},\ and\ \citenamefont {Meir}}]{Golub2006}%
  \BibitemOpen
  \bibfield  {author} {\bibinfo {author} {\bibfnamefont {A.}~\bibnamefont
  {Golub}}, \bibinfo {author} {\bibfnamefont {T.}~\bibnamefont {Aono}}, \ and\
  \bibinfo {author} {\bibfnamefont {Y.}~\bibnamefont {Meir}},\ }\href {\doibase
  10.1103/PhysRevLett.97.186801} {\bibfield  {journal} {\bibinfo  {journal}
  {Phys. Rev. Lett.}\ }\textbf {\bibinfo {volume} {97}},\ \bibinfo {eid}
  {186801} (\bibinfo {year} {2006})}\BibitemShut {NoStop}%
\bibitem [{\citenamefont {Rejec}\ and\ \citenamefont {Meir}(2006)}]{Rejec2006}%
  \BibitemOpen
  \bibfield  {author} {\bibinfo {author} {\bibfnamefont {T.}~\bibnamefont
  {Rejec}}\ and\ \bibinfo {author} {\bibfnamefont {Y.}~\bibnamefont {Meir}},\
  }\href {\doibase doi: 10.1038/nature05054} {\bibfield  {journal} {\bibinfo
  {journal} {Nature}\ }\textbf {\bibinfo {volume} {442}},\ \bibinfo {pages}
  {900} (\bibinfo {year} {2006})}\BibitemShut {NoStop}%
\bibitem [{\citenamefont {Chen}\ \emph {et~al.}(2008)\citenamefont {Chen},
  \citenamefont {Graham}, \citenamefont {Pepper}, \citenamefont {Farrer},\ and\
  \citenamefont {Ritchie}}]{Chen2008}%
  \BibitemOpen
  \bibfield  {author} {\bibinfo {author} {\bibfnamefont {T.-M.}\ \bibnamefont
  {Chen}}, \bibinfo {author} {\bibfnamefont {A.~C.}\ \bibnamefont {Graham}},
  \bibinfo {author} {\bibfnamefont {M.}~\bibnamefont {Pepper}}, \bibinfo
  {author} {\bibfnamefont {I.}~\bibnamefont {Farrer}}, \ and\ \bibinfo {author}
  {\bibfnamefont {D.~A.}\ \bibnamefont {Ritchie}},\ }\href {\doibase
  10.1063/1.2963478} {\bibfield  {journal} {\bibinfo  {journal} {Applied
  Physics Letters}\ }\textbf {\bibinfo {volume} {93}},\ \bibinfo {eid} {032102}
  (\bibinfo {year} {2008})}\BibitemShut {NoStop}%
\bibitem [{\citenamefont {Chen}\ \emph {et~al.}(2010)\citenamefont {Chen},
  \citenamefont {Graham}, \citenamefont {Pepper}, \citenamefont {Farrer},
  \citenamefont {Anderson}, \citenamefont {Jones},\ and\ \citenamefont
  {Ritchie}}]{Chen2010a}%
  \BibitemOpen
  \bibfield  {author} {\bibinfo {author} {\bibfnamefont {T.-M.}\ \bibnamefont
  {Chen}}, \bibinfo {author} {\bibfnamefont {A.~C.}\ \bibnamefont {Graham}},
  \bibinfo {author} {\bibfnamefont {M.}~\bibnamefont {Pepper}}, \bibinfo
  {author} {\bibfnamefont {I.}~\bibnamefont {Farrer}}, \bibinfo {author}
  {\bibfnamefont {D.}~\bibnamefont {Anderson}}, \bibinfo {author}
  {\bibfnamefont {G.~A.~C.}\ \bibnamefont {Jones}}, \ and\ \bibinfo {author}
  {\bibfnamefont {D.~A.}\ \bibnamefont {Ritchie}},\ }\href@noop {} {\bibfield
  {journal} {\bibinfo  {journal} {Nano}\ }\textbf {\bibinfo {volume} {10}},\
  \bibinfo {pages} {2330} (\bibinfo {year} {2010})}\BibitemShut {NoStop}%
\bibitem [{\citenamefont {Chen}\ \emph {et~al.}(2012)\citenamefont {Chen},
  \citenamefont {Pepper}, \citenamefont {Farrer}, \citenamefont {Jones},\ and\
  \citenamefont {Ritchie}}]{Chen2012}%
  \BibitemOpen
  \bibfield  {author} {\bibinfo {author} {\bibfnamefont {T.-M.}\ \bibnamefont
  {Chen}}, \bibinfo {author} {\bibfnamefont {M.}~\bibnamefont {Pepper}},
  \bibinfo {author} {\bibfnamefont {I.}~\bibnamefont {Farrer}}, \bibinfo
  {author} {\bibfnamefont {G.~A.~C.}\ \bibnamefont {Jones}}, \ and\ \bibinfo
  {author} {\bibfnamefont {D.~A.}\ \bibnamefont {Ritchie}},\ }\href {\doibase
  10.1103/PhysRevLett.109.177202} {\bibfield  {journal} {\bibinfo  {journal}
  {Phys. Rev. Lett.}\ }\textbf {\bibinfo {volume} {109}},\ \bibinfo {pages}
  {177202} (\bibinfo {year} {2012})}\BibitemShut {NoStop}%
\bibitem [{\citenamefont {Potok}\ \emph {et~al.}(2002)\citenamefont {Potok},
  \citenamefont {Folk}, \citenamefont {Marcus},\ and\ \citenamefont
  {Umansky}}]{Potok2002}%
  \BibitemOpen
  \bibfield  {author} {\bibinfo {author} {\bibfnamefont {R.~M.}\ \bibnamefont
  {Potok}}, \bibinfo {author} {\bibfnamefont {J.~A.}\ \bibnamefont {Folk}},
  \bibinfo {author} {\bibfnamefont {C.~M.}\ \bibnamefont {Marcus}}, \ and\
  \bibinfo {author} {\bibfnamefont {V.}~\bibnamefont {Umansky}},\ }\href
  {\doibase 10.1103/PhysRevLett.89.266602} {\bibfield  {journal} {\bibinfo
  {journal} {Phys. Rev. Lett.}\ }\textbf {\bibinfo {volume} {89}},\ \bibinfo
  {pages} {266602} (\bibinfo {year} {2002})}\BibitemShut {NoStop}%
\bibitem [{\citenamefont {Wang}\ and\ \citenamefont
  {Berggren}(1996)}]{Wang1996}%
  \BibitemOpen
  \bibfield  {author} {\bibinfo {author} {\bibfnamefont {C.-K.}\ \bibnamefont
  {Wang}}\ and\ \bibinfo {author} {\bibfnamefont {K.-F.}\ \bibnamefont
  {Berggren}},\ }\href {\doibase 10.1103/PhysRevB.54.R14257} {\bibfield
  {journal} {\bibinfo  {journal} {Phys. Rev. B}\ }\textbf {\bibinfo {volume}
  {54}},\ \bibinfo {pages} {R14257} (\bibinfo {year} {1996})}\BibitemShut
  {NoStop}%
\bibitem [{\citenamefont {Wang}\ and\ \citenamefont
  {Berggren}(1998)}]{Wang1998}%
  \BibitemOpen
  \bibfield  {author} {\bibinfo {author} {\bibfnamefont {C.-K.}\ \bibnamefont
  {Wang}}\ and\ \bibinfo {author} {\bibfnamefont {K.-F.}\ \bibnamefont
  {Berggren}},\ }\href {\doibase 10.1103/PhysRevB.57.4552} {\bibfield
  {journal} {\bibinfo  {journal} {Phys. Rev. B}\ }\textbf {\bibinfo {volume}
  {57}},\ \bibinfo {pages} {4552} (\bibinfo {year} {1998})}\BibitemShut
  {NoStop}%
\bibitem [{\citenamefont {Micolich}(2011)}]{Micolich2011}%
  \BibitemOpen
  \bibfield  {author} {\bibinfo {author} {\bibfnamefont {A.~P.}\ \bibnamefont
  {Micolich}},\ }\href {http://stacks.iop.org/0953-8984/23/i=44/a=443201}
  {\bibfield  {journal} {\bibinfo  {journal} {Journal of Physics: Condensed
  Matter}\ }\textbf {\bibinfo {volume} {23}},\ \bibinfo {pages} {443201}
  (\bibinfo {year} {2011})}\BibitemShut {NoStop}%
\bibitem [{\citenamefont {Komijani}\ \emph {et~al.}(2010)\citenamefont
  {Komijani}, \citenamefont {Csontos}, \citenamefont {Shorubalko},
  \citenamefont {Ihn}, \citenamefont {Ensslin}, \citenamefont {Meir},
  \citenamefont {Reuter},\ and\ \citenamefont {Wieck}}]{Komijani2009}%
  \BibitemOpen
  \bibfield  {author} {\bibinfo {author} {\bibfnamefont {Y.}~\bibnamefont
  {Komijani}}, \bibinfo {author} {\bibfnamefont {M.}~\bibnamefont {Csontos}},
  \bibinfo {author} {\bibfnamefont {I.}~\bibnamefont {Shorubalko}}, \bibinfo
  {author} {\bibfnamefont {T.}~\bibnamefont {Ihn}}, \bibinfo {author}
  {\bibfnamefont {K.}~\bibnamefont {Ensslin}}, \bibinfo {author} {\bibfnamefont
  {Y.}~\bibnamefont {Meir}}, \bibinfo {author} {\bibfnamefont {D.}~\bibnamefont
  {Reuter}}, \ and\ \bibinfo {author} {\bibfnamefont {A.~D.}\ \bibnamefont
  {Wieck}},\ }\href {\doibase doi:10.1209/0295-5075/91/67010} {\bibfield
  {journal} {\bibinfo  {journal} {Eur. Phys. Lett.}\ }\textbf {\bibinfo
  {volume} {91}},\ \bibinfo {pages} {67010} (\bibinfo {year}
  {2010})}\BibitemShut {NoStop}%
\bibitem [{\citenamefont {Chung}\ \emph {et~al.}(2007)\citenamefont {Chung},
  \citenamefont {Jo}, \citenamefont {Chang}, \citenamefont {Lee}, \citenamefont
  {Zaffalon}, \citenamefont {Umansky},\ and\ \citenamefont
  {Heiblum}}]{Chung2007}%
  \BibitemOpen
  \bibfield  {author} {\bibinfo {author} {\bibfnamefont {Y.}~\bibnamefont
  {Chung}}, \bibinfo {author} {\bibfnamefont {S.}~\bibnamefont {Jo}}, \bibinfo
  {author} {\bibfnamefont {D.-I.}\ \bibnamefont {Chang}}, \bibinfo {author}
  {\bibfnamefont {H.-J.}\ \bibnamefont {Lee}}, \bibinfo {author} {\bibfnamefont
  {M.}~\bibnamefont {Zaffalon}}, \bibinfo {author} {\bibfnamefont
  {V.}~\bibnamefont {Umansky}}, \ and\ \bibinfo {author} {\bibfnamefont
  {M.}~\bibnamefont {Heiblum}},\ }\href {\doibase 10.1103/PhysRevB.76.035316}
  {\bibfield  {journal} {\bibinfo  {journal} {Phys. Rev. B}\ }\textbf {\bibinfo
  {volume} {76}},\ \bibinfo {eid} {035316} (\bibinfo {year}
  {2007})}\BibitemShut {NoStop}%
\bibitem [{\citenamefont {Bauer}\ \emph {et~al.}(2013)\citenamefont {Bauer},
  \citenamefont {Heyder}, \citenamefont {Schubert}, \citenamefont {Borowsky},
  \citenamefont {Taubert}, \citenamefont {Bruognolo}, \citenamefont {Schuh},
  \citenamefont {Wegscheider}, \citenamefont {von Delft},\ and\ \citenamefont
  {Ludwig}}]{Bauer2013}%
  \BibitemOpen
  \bibfield  {author} {\bibinfo {author} {\bibfnamefont {F.}~\bibnamefont
  {Bauer}}, \bibinfo {author} {\bibfnamefont {J.}~\bibnamefont {Heyder}},
  \bibinfo {author} {\bibfnamefont {E.}~\bibnamefont {Schubert}}, \bibinfo
  {author} {\bibfnamefont {D.}~\bibnamefont {Borowsky}}, \bibinfo {author}
  {\bibfnamefont {D.}~\bibnamefont {Taubert}}, \bibinfo {author} {\bibfnamefont
  {B.}~\bibnamefont {Bruognolo}}, \bibinfo {author} {\bibfnamefont
  {D.}~\bibnamefont {Schuh}}, \bibinfo {author} {\bibfnamefont
  {W.}~\bibnamefont {Wegscheider}}, \bibinfo {author} {\bibfnamefont
  {J.}~\bibnamefont {von Delft}}, \ and\ \bibinfo {author} {\bibfnamefont
  {S.}~\bibnamefont {Ludwig}},\ }\href {\doibase
  http://dx.doi.org/10.1038/nature12421 10.1038/nature12421} {\bibfield
  {journal} {\bibinfo  {journal} {Nature}\ }\textbf {\bibinfo {volume} {501}},\
  \bibinfo {pages} {73} (\bibinfo {year} {2013})}\BibitemShut {NoStop}%
\bibitem [{\citenamefont {Heyder}\ \emph
  {et~al.}(2015{\natexlab{a}})\citenamefont {Heyder}, \citenamefont {Bauer},
  \citenamefont {Schimmel},\ and\ \citenamefont {von Delft}}]{Heyder2016}%
  \BibitemOpen
  \bibfield  {author} {\bibinfo {author} {\bibfnamefont {J.}~\bibnamefont
  {Heyder}}, \bibinfo {author} {\bibfnamefont {F.}~\bibnamefont {Bauer}},
  \bibinfo {author} {\bibfnamefont {D.}~\bibnamefont {Schimmel}}, \ and\
  \bibinfo {author} {\bibfnamefont {J.}~\bibnamefont {von Delft}},\ }\href@noop
  {} {\bibfield  {journal} {\bibinfo  {journal} {to be published}\ } (\bibinfo
  {year} {2015}{\natexlab{a}})}\BibitemShut {NoStop}%
\bibitem [{\citenamefont {Bauer}\ \emph {et~al.}(2014)\citenamefont {Bauer},
  \citenamefont {Heyder},\ and\ \citenamefont {von Delft}}]{Bauer2014}%
  \BibitemOpen
  \bibfield  {author} {\bibinfo {author} {\bibfnamefont {F.}~\bibnamefont
  {Bauer}}, \bibinfo {author} {\bibfnamefont {J.}~\bibnamefont {Heyder}}, \
  and\ \bibinfo {author} {\bibfnamefont {J.}~\bibnamefont {von Delft}},\ }\href
  {\doibase 10.1103/PhysRevB.89.045128} {\bibfield  {journal} {\bibinfo
  {journal} {Phys. Rev. B}\ }\textbf {\bibinfo {volume} {89}},\ \bibinfo
  {pages} {045128} (\bibinfo {year} {2014})}\BibitemShut {NoStop}%
\bibitem [{\citenamefont {Heyder}\ \emph
  {et~al.}(2015{\natexlab{b}})\citenamefont {Heyder}, \citenamefont {Bauer},
  \citenamefont {Schubert}, \citenamefont {Borowsky}, \citenamefont {Schuh},
  \citenamefont {Wegscheider}, \citenamefont {von Delft},\ and\ \citenamefont
  {Ludwig}}]{Heyder2015}%
  \BibitemOpen
  \bibfield  {author} {\bibinfo {author} {\bibfnamefont {J.}~\bibnamefont
  {Heyder}}, \bibinfo {author} {\bibfnamefont {F.}~\bibnamefont {Bauer}},
  \bibinfo {author} {\bibfnamefont {E.}~\bibnamefont {Schubert}}, \bibinfo
  {author} {\bibfnamefont {D.}~\bibnamefont {Borowsky}}, \bibinfo {author}
  {\bibfnamefont {D.}~\bibnamefont {Schuh}}, \bibinfo {author} {\bibfnamefont
  {W.}~\bibnamefont {Wegscheider}}, \bibinfo {author} {\bibfnamefont
  {J.}~\bibnamefont {von Delft}}, \ and\ \bibinfo {author} {\bibfnamefont
  {S.}~\bibnamefont {Ludwig}},\ }\href {\doibase 10.1103/PhysRevB.92.195401}
  {\bibfield  {journal} {\bibinfo  {journal} {Phys. Rev. B}\ }\textbf {\bibinfo
  {volume} {92}},\ \bibinfo {pages} {195401} (\bibinfo {year}
  {2015}{\natexlab{b}})},\ \bibinfo {note} {arXiv:1409.3415
  [cond-mat.str-el]}\BibitemShut {NoStop}%
\bibitem [{\citenamefont {Jakobs}\ \emph
  {et~al.}(2010{\natexlab{a}})\citenamefont {Jakobs}, \citenamefont
  {Pletyukhov},\ and\ \citenamefont {Schoeller}}]{Jakobs2010}%
  \BibitemOpen
  \bibfield  {author} {\bibinfo {author} {\bibfnamefont {S.~G.}\ \bibnamefont
  {Jakobs}}, \bibinfo {author} {\bibfnamefont {M.}~\bibnamefont {Pletyukhov}},
  \ and\ \bibinfo {author} {\bibfnamefont {H.}~\bibnamefont {Schoeller}},\
  }\href {\doibase 10.1103/PhysRevB.81.195109} {\bibfield  {journal} {\bibinfo
  {journal} {Phys. Rev. B}\ }\textbf {\bibinfo {volume} {81}},\ \bibinfo
  {pages} {195109} (\bibinfo {year} {2010}{\natexlab{a}})}\BibitemShut
  {NoStop}%
\bibitem [{\citenamefont {Jakobs}\ \emph
  {et~al.}(2010{\natexlab{b}})\citenamefont {Jakobs}, \citenamefont
  {Pletyukhov},\ and\ \citenamefont {Schoeller}}]{Jakobs2010-2}%
  \BibitemOpen
  \bibfield  {author} {\bibinfo {author} {\bibfnamefont {S.~G.}\ \bibnamefont
  {Jakobs}}, \bibinfo {author} {\bibfnamefont {M.}~\bibnamefont {Pletyukhov}},
  \ and\ \bibinfo {author} {\bibfnamefont {H.}~\bibnamefont {Schoeller}},\
  }\href {http://stacks.iop.org/1751-8121/43/i=10/a=103001} {\bibfield
  {journal} {\bibinfo  {journal} {J. Phys. A: Math. and Theor.}\ }\textbf
  {\bibinfo {volume} {43}},\ \bibinfo {pages} {103001} (\bibinfo {year}
  {2010}{\natexlab{b}})}\BibitemShut {NoStop}%
\bibitem [{\citenamefont {Karrasch}\ \emph {et~al.}(2006)\citenamefont
  {Karrasch}, \citenamefont {Enss},\ and\ \citenamefont
  {Meden}}]{Karrasch2006a}%
  \BibitemOpen
  \bibfield  {author} {\bibinfo {author} {\bibfnamefont {C.}~\bibnamefont
  {Karrasch}}, \bibinfo {author} {\bibfnamefont {T.}~\bibnamefont {Enss}}, \
  and\ \bibinfo {author} {\bibfnamefont {V.}~\bibnamefont {Meden}},\ }\href
  {\doibase 10.1103/PhysRevB.73.235337} {\bibfield  {journal} {\bibinfo
  {journal} {Phys. Rev. B}\ }\textbf {\bibinfo {volume} {73}},\ \bibinfo {eid}
  {235337} (\bibinfo {year} {2006})}\BibitemShut {NoStop}%
\bibitem [{\citenamefont {Metzner}\ \emph {et~al.}(2012)\citenamefont
  {Metzner}, \citenamefont {Salmhofer}, \citenamefont {Honerkamp},
  \citenamefont {Meden},\ and\ \citenamefont {Sch\"onhammer}}]{Metzner2012}%
  \BibitemOpen
  \bibfield  {author} {\bibinfo {author} {\bibfnamefont {W.}~\bibnamefont
  {Metzner}}, \bibinfo {author} {\bibfnamefont {M.}~\bibnamefont {Salmhofer}},
  \bibinfo {author} {\bibfnamefont {C.}~\bibnamefont {Honerkamp}}, \bibinfo
  {author} {\bibfnamefont {V.}~\bibnamefont {Meden}}, \ and\ \bibinfo {author}
  {\bibfnamefont {K.}~\bibnamefont {Sch\"onhammer}},\ }\href@noop {} {\bibfield
   {journal} {\bibinfo  {journal} {Rev. Mod. Phys.}\ }\textbf {\bibinfo
  {volume} {84}},\ \bibinfo {pages} {299} (\bibinfo {year} {2012})}\BibitemShut
  {NoStop}%
\bibitem [{sup()}]{supplement}%
  \BibitemOpen
  \href@noop {} {\bibinfo  {journal} {See Supplemental Material at [url] for
  the technical details used in our calculation, which includes
  Refs.~\cite{Schollwoeck2005,White1992,White1993,Schollwoeck2011,Vekic_PRB_1996,
  Vekic_PRL_1993, Vidal04, *Daley04, *WhiteFeiguin04, White_PRB_2008_LinPre,
  Barthel_PRB_2009, weichselbaum12a}}\ }\BibitemShut {NoStop}%
\bibitem [{\citenamefont {Ihnatsenka}\ and\ \citenamefont
  {Zozoulenko}(2009)}]{PhysRevB.79.235313}%
  \BibitemOpen
\bibfield  {journal} {  }\bibfield  {author} {\bibinfo {author} {\bibfnamefont
  {S.}~\bibnamefont {Ihnatsenka}}\ and\ \bibinfo {author} {\bibfnamefont
  {I.~V.}\ \bibnamefont {Zozoulenko}},\ }\href {\doibase
  10.1103/PhysRevB.79.235313} {\bibfield  {journal} {\bibinfo  {journal} {Phys.
  Rev. B}\ }\textbf {\bibinfo {volume} {79}},\ \bibinfo {pages} {235313}
  (\bibinfo {year} {2009})}\BibitemShut {NoStop}%
\bibitem [{\citenamefont {S\'anchez}\ and\ \citenamefont
  {Leburton}(2013)}]{PhysRevB.88.075305}%
  \BibitemOpen
  \bibfield  {author} {\bibinfo {author} {\bibfnamefont {A.~X.}\ \bibnamefont
  {S\'anchez}}\ and\ \bibinfo {author} {\bibfnamefont {J.-P.}\ \bibnamefont
  {Leburton}},\ }\href {\doibase 10.1103/PhysRevB.88.075305} {\bibfield
  {journal} {\bibinfo  {journal} {Phys. Rev. B}\ }\textbf {\bibinfo {volume}
  {88}},\ \bibinfo {pages} {075305} (\bibinfo {year} {2013})}\BibitemShut
  {NoStop}%
\bibitem [{\citenamefont {Aryanpour}\ and\ \citenamefont
  {Han}(2009)}]{Aryanpour2009}%
  \BibitemOpen
  \bibfield  {author} {\bibinfo {author} {\bibfnamefont {K.}~\bibnamefont
  {Aryanpour}}\ and\ \bibinfo {author} {\bibfnamefont {J.~E.}\ \bibnamefont
  {Han}},\ }\href {\doibase 10.1103/PhysRevLett.102.056805} {\bibfield
  {journal} {\bibinfo  {journal} {Phys. Rev. Lett.}\ }\textbf {\bibinfo
  {volume} {102}},\ \bibinfo {pages} {056805} (\bibinfo {year}
  {2009})}\BibitemShut {NoStop}%
\bibitem [{\citenamefont {Wigner}(1955)}]{Wigner1955}%
  \BibitemOpen
  \bibfield  {author} {\bibinfo {author} {\bibfnamefont {E.~P.}\ \bibnamefont
  {Wigner}},\ }\href {\doibase 10.1103/PhysRev.98.145} {\bibfield  {journal}
  {\bibinfo  {journal} {Phys. Rev.}\ }\textbf {\bibinfo {volume} {98}},\
  \bibinfo {pages} {145} (\bibinfo {year} {1955})}\BibitemShut {NoStop}%
\bibitem [{\citenamefont {Schollw\"ock}(2005)}]{Schollwoeck2005}%
  \BibitemOpen
  \bibfield  {author} {\bibinfo {author} {\bibfnamefont {U.}~\bibnamefont
  {Schollw\"ock}},\ }\href {\doibase 10.1103/RevModPhys.77.259} {\bibfield
  {journal} {\bibinfo  {journal} {Rev. Mod. Phys.}\ }\textbf {\bibinfo {volume}
  {77}},\ \bibinfo {pages} {259} (\bibinfo {year} {2005})}\BibitemShut
  {NoStop}%
\bibitem [{\citenamefont {White}(1992)}]{White1992}%
  \BibitemOpen
  \bibfield  {author} {\bibinfo {author} {\bibfnamefont {S.~R.}\ \bibnamefont
  {White}},\ }\href {\doibase 10.1103/PhysRevLett.69.2863} {\bibfield
  {journal} {\bibinfo  {journal} {Phys. Rev. Lett.}\ }\textbf {\bibinfo
  {volume} {69}},\ \bibinfo {pages} {2863} (\bibinfo {year}
  {1992})}\BibitemShut {NoStop}%
\bibitem [{\citenamefont {White}(1993)}]{White1993}%
  \BibitemOpen
  \bibfield  {author} {\bibinfo {author} {\bibfnamefont {S.~R.}\ \bibnamefont
  {White}},\ }\href {\doibase 10.1103/PhysRevB.48.10345} {\bibfield  {journal}
  {\bibinfo  {journal} {Phys. Rev. B}\ }\textbf {\bibinfo {volume} {48}},\
  \bibinfo {pages} {10345} (\bibinfo {year} {1993})}\BibitemShut {NoStop}%
\bibitem [{\citenamefont {Schollw\"ock}(2011)}]{Schollwoeck2011}%
  \BibitemOpen
  \bibfield  {author} {\bibinfo {author} {\bibfnamefont {U.}~\bibnamefont
  {Schollw\"ock}},\ }\href {\doibase 10.1016/j.aop.2010.09.012} {\bibfield
  {journal} {\bibinfo  {journal} {Ann. Phys.}\ }\textbf {\bibinfo {volume}
  {326}},\ \bibinfo {pages} {96 } (\bibinfo {year} {2011})}\BibitemShut
  {NoStop}%
\bibitem [{\citenamefont {Veki\ifmmode~\acute{c}\else \'{c}\fi{}}\ and\
  \citenamefont {White}(1996)}]{Vekic_PRB_1996}%
  \BibitemOpen
  \bibfield  {author} {\bibinfo {author} {\bibfnamefont {M.}~\bibnamefont
  {Veki\ifmmode~\acute{c}\else \'{c}\fi{}}}\ and\ \bibinfo {author}
  {\bibfnamefont {S.~R.}\ \bibnamefont {White}},\ }\href {\doibase
  10.1103/PhysRevB.53.14552} {\bibfield  {journal} {\bibinfo  {journal} {Phys.
  Rev. B}\ }\textbf {\bibinfo {volume} {53}},\ \bibinfo {pages} {14552}
  (\bibinfo {year} {1996})}\BibitemShut {NoStop}%
\bibitem [{\citenamefont {Veki\ifmmode~\acute{c}\else \'{c}\fi{}}\ and\
  \citenamefont {White}(1993)}]{Vekic_PRL_1993}%
  \BibitemOpen
  \bibfield  {author} {\bibinfo {author} {\bibfnamefont {M.}~\bibnamefont
  {Veki\ifmmode~\acute{c}\else \'{c}\fi{}}}\ and\ \bibinfo {author}
  {\bibfnamefont {S.~R.}\ \bibnamefont {White}},\ }\href {\doibase
  10.1103/PhysRevLett.71.4283} {\bibfield  {journal} {\bibinfo  {journal}
  {Phys. Rev. Lett.}\ }\textbf {\bibinfo {volume} {71}},\ \bibinfo {pages}
  {4283} (\bibinfo {year} {1993})}\BibitemShut {NoStop}%
\bibitem [{\citenamefont {Vidal}(2004)}]{Vidal04}%
  \BibitemOpen
  \bibfield  {author} {\bibinfo {author} {\bibfnamefont {G.}~\bibnamefont
  {Vidal}},\ }\href {\doibase 10.1103/PhysRevLett.93.040502} {\bibfield
  {journal} {\bibinfo  {journal} {Phys. Rev. Lett.}\ }\textbf {\bibinfo
  {volume} {93}},\ \bibinfo {pages} {040502} (\bibinfo {year}
  {2004})}\BibitemShut {NoStop}%
\bibitem [{\citenamefont {Daley}\ \emph {et~al.}(2004)\citenamefont {Daley},
  \citenamefont {Kollath}, \citenamefont {Schollw\"ock},\ and\ \citenamefont
  {Vidal}}]{Daley04}%
  \BibitemOpen
  \bibfield  {author} {\bibinfo {author} {\bibfnamefont {A.~J.}\ \bibnamefont
  {Daley}}, \bibinfo {author} {\bibfnamefont {C.}~\bibnamefont {Kollath}},
  \bibinfo {author} {\bibfnamefont {U.}~\bibnamefont {Schollw\"ock}}, \ and\
  \bibinfo {author} {\bibfnamefont {G.}~\bibnamefont {Vidal}},\ }\href
  {http://stacks.iop.org/1742-5468/2004/i=04/a=P04005} {\bibfield  {journal}
  {\bibinfo  {journal} {Journal of Statistical Mechanics: Theory and
  Experiment}\ }\textbf {\bibinfo {volume} {2004}},\ \bibinfo {pages} {P04005}
  (\bibinfo {year} {2004})}\BibitemShut {NoStop}%
\bibitem [{\citenamefont {White}\ and\ \citenamefont
  {Feiguin}(2004)}]{WhiteFeiguin04}%
  \BibitemOpen
  \bibfield  {author} {\bibinfo {author} {\bibfnamefont {S.~R.}\ \bibnamefont
  {White}}\ and\ \bibinfo {author} {\bibfnamefont {A.~E.}\ \bibnamefont
  {Feiguin}},\ }\href {\doibase 10.1103/PhysRevLett.93.076401} {\bibfield
  {journal} {\bibinfo  {journal} {Phys. Rev. Lett.}\ }\textbf {\bibinfo
  {volume} {93}},\ \bibinfo {pages} {076401} (\bibinfo {year}
  {2004})}\BibitemShut {NoStop}%
\bibitem [{\citenamefont {White}\ and\ \citenamefont
  {Affleck}(2008)}]{White_PRB_2008_LinPre}%
  \BibitemOpen
  \bibfield  {author} {\bibinfo {author} {\bibfnamefont {S.~R.}\ \bibnamefont
  {White}}\ and\ \bibinfo {author} {\bibfnamefont {I.}~\bibnamefont
  {Affleck}},\ }\href {\doibase 10.1103/PhysRevB.77.134437} {\bibfield
  {journal} {\bibinfo  {journal} {Phys. Rev. B}\ }\textbf {\bibinfo {volume}
  {77}},\ \bibinfo {pages} {134437} (\bibinfo {year} {2008})}\BibitemShut
  {NoStop}%
\bibitem [{\citenamefont {Barthel}\ \emph {et~al.}(2009)\citenamefont
  {Barthel}, \citenamefont {Schollw\"ock},\ and\ \citenamefont
  {White}}]{Barthel_PRB_2009}%
  \BibitemOpen
  \bibfield  {author} {\bibinfo {author} {\bibfnamefont {T.}~\bibnamefont
  {Barthel}}, \bibinfo {author} {\bibfnamefont {U.}~\bibnamefont
  {Schollw\"ock}}, \ and\ \bibinfo {author} {\bibfnamefont {S.~R.}\
  \bibnamefont {White}},\ }\href {\doibase 10.1103/PhysRevB.79.245101}
  {\bibfield  {journal} {\bibinfo  {journal} {Phys. Rev. B}\ }\textbf {\bibinfo
  {volume} {79}},\ \bibinfo {pages} {245101} (\bibinfo {year}
  {2009})}\BibitemShut {NoStop}%
\bibitem [{\citenamefont {Weichselbaum}(2012)}]{weichselbaum12a}%
  \BibitemOpen
  \bibfield  {author} {\bibinfo {author} {\bibfnamefont {A.}~\bibnamefont
  {Weichselbaum}},\ }\href {\doibase 10.1016/j.aop.2012.07.009} {\bibfield
  {journal} {\bibinfo  {journal} {Ann. Phys.}\ }\textbf {\bibinfo {volume}
  {327}},\ \bibinfo {pages} {2972 } (\bibinfo {year} {2012})}\BibitemShut
  {NoStop}%
\end{thebibliography}%

\newpage
\begin{center}
\textbf{\large Supplementary material}
\end{center}

\setcounter{equation}{0}
\setcounter{figure}{0}
\renewcommand{\theequation}{S\arabic{equation}}
\renewcommand{\thefigure}{S\arabic{figure}}
\renewcommand{\thesection}{{S-\arabic{section}}}
\renewcommand{\thesubsection}{{S-\arabic{section}.\arabic{subsection}}}

This supplement consists of two parts. In the first,
we give the technical details on the model, the fRG-
flow equations and the numerics involved. 
We also argue that
the characteristic frequency for spin fluctuations,
$\omega_{\rm spin}$, is governed by the distance between the chemical
potential and the effective lower band edge, 
$\omega_{\rm spin} \simeq \mu - \omega_{\rm max}$.
In the second part, we report on DMRG calculations
of the LDOS that we have performed to as an independent
check of our fRG predictions. We find good qualitative
agreement between both methods.

\section{S-I. Model}
We use a modified version of Model II of Ref.~\cite{Bauer2013}: In the central region, described by $N=2N'+1$ sites, with $i=-N', \dots, N'$, the on-site potential is zero, and the hopping elements vary from site to site according to
\begin{equation}
	\tau_j = \tau-\tfrac 12 \tilde \Vg \exp \left(-\frac{x_j^2}{1-x_j^2} \right); \; x_j = \frac{2j+1}{N-1},
\label{eq:hopping}
\end{equation}
where $j$ runs from $-N'$ to $N'-1$.
The on-site interaction in the central region is given by
\begin{equation}
\label{eq:interaction}
	U_i = U_0 \exp \left(-\frac{l_i^6}{1-l_i^2} \right); \; l_i = \frac{i}{N'+\tfrac 12},
\end{equation}
The hopping and interaction Eqs.~\eqref{eq:hopping},\eqref{eq:interaction} lead to a Hamiltonian 
\begin{align}
\label{eq:define_H_general}
	\mathcal H &= -\sum_{\sigma, i} \tau_{i} \left( c^\dagger_{i+1,\sigma} c_{i,\sigma} + {\rm h.c.} \right) + \sum_{i} \left( U_i c^\dagger_{i\uparrow} c_{i\uparrow} c^\dagger_{i\downarrow} c_{i\downarrow} \right), \nonumber \\
	           &=: \sum_{\sigma, i,j} \left( \tilde H^\sigma_{ji} c^\dagger_{j,\sigma} c_{i,\sigma} + {\rm h.c.} \right) + \sum_{i} \left( U_i c^\dagger_{i\uparrow} c_{i\uparrow} c^\dagger_{i\downarrow} c_{i\downarrow} \right),
\end{align}
where we use the tilde to indicate that the indices of the Hamiltonian matrix $\tilde H^\sigma$ run over $\mathbbm Z$.
$\tilde H^\sigma_{ij}$ is invariant under transposition and parity ${\cal P}$, which we implement as $ {\cal P}: i \mapsto -i$. We will explicitly assume the presence of these symmetries in the following.
Note that for our description of the central region, the effect of the tight-binding leads with hopping $\tau$ coupling to sites $-N'$ and $N'$ is fully included in the self-energy contribution
\begin{align}
\label{eq:selfenergy_lead}
	{\Sigma^{R}_{\rm lead}}_{ij} (\omega) =& (\delta_{i,-N'} \delta_{j,-N'} + \delta_{i,N'} \delta_{j,N'}) \nonumber \\
	& \times
	\begin{cases}
		\frac{\omega}{2} \left( 1 - \sqrt{1-\left( \frac{2 \tau}{\omega} \right)^2 } \right), \, \vert \omega \vert > 2 \tau \\
		\frac{\omega}{2} - i \tau \sqrt{1 - \left( \frac{\omega}{2 \tau} \right)^2},          \, \vert \omega \vert < 2 \tau,
	\end{cases} \\
	{\Sigma^{K}_{\rm lead}}_{ij} (\omega) =& (1-2n_F(\omega))({\Sigma^R_{\rm lead}}_{ij}-{\Sigma^A_{\rm lead}}_{ij}).
\end{align}
Here, the superscript $R$($K$,$A$) denotes the retarded (Keldysh, advanced) component of the self energy and $n_F$ is the Fermi distribution function.

As stated in the main text, we use $U_0 = 0.7 \tau$ and $\tilde \Vg \in [0.44, 0.58] \tau$.

\section{S-II. Keldysh fRG}
\label{sec:fRG_flow_equations}
The model is solved by employing the functional renormalization group (fRG) \cite{Jakobs2010,Jakobs2010-2,Karrasch2006a,Metzner2012} on the Keldysh-contour to obtain real-frequency information. The flow is truncated perturbatively, i.e. we set the three-particle vertex (and all higher vertices) to zero during the flow and approximate the two-particle vertex by the three usual channels ($P$, $X$, and $D$) \cite{Bauer2013,Jakobs2010}, assuming a local and static inter-channel mixing (coupled-ladder-approximation). The computation is then exact to second order in the interaction. It may be viewed as extension of the flow used in Ref.~\cite{Jakobs2010-2} to multiple sites (neglecting the $D^{\sigma \bar \sigma}$-channel, which in our case is of order $U_0^3$) or an extension of the flow used in Ref.~\cite{Bauer2013} to real frequencies. As flow parameter we use an artificial, on-site broadening of the spectrum (c.f.~Eq.~\eqref{eq:green_function_with_flow_param}, and Ref.~\cite{Jakobs2010}).
This flow parameter respects fluctuation-dissipation theorems, so that in equilibrium it is unnecessary to compute the Keldysh components of the self energy ($\Sigma^K$) and the channels ($b^P$, $b^X$, $b^D$).
The conventions on the Keldysh-contour used are those of Ref.~\cite{Jakobs2010-2}, with the difference that after the Keldysh rotation we use the labels c(lassical) and q(uantum), instead of $2$ and $1$. In particular, this means that the Keldysh rotation used for the fermions is the same as the one usually used for bosons.
We use $\sigma = \uparrow, \downarrow$ to denote spin, and $\bar \sigma$ to denote the spin opposite to $\sigma$. Letters from the middle of the roman alphabet ($i$,$j$) refer to spatial sites, while letters from the beginning of the Greek alphabet ($\alpha$, $\beta$) refer to the Keldysh indices.

\subsection{A. The Single-Scale Propagator}
The flow parameter is determined by the bare retarded Green's function
\begin{equation}
\label{eq:green_function_with_flow_param}
	\tilde G^R_{0,\Lambda, \sigma} (\omega) = \frac{1}{\omega {\mathbbm 1}- \tilde H^\sigma + i \left( \tfrac 12 \Lambda \right) {\mathbbm 1}},
\end{equation}
where $\tilde H^\sigma$ is the non-interacting Hamiltonian matrix extracted from Eq.~\eqref{eq:define_H_general}. $\Lambda$ is the flow parameter, ranging from $\infty$ (start of flow) to $0$ (end of flow). $\mathbbm 1$ is the unit matrix in the space of the sites, which we will omit from now on. Once the leads have been projected out, we drop the tilde on the restricted Hamiltonian matrix $H^\sigma$ and the spatial indices then only run from $-N'$ to $N'$. We use the artificial on-site broadening for all sites (including the leads) to avoid artifacts at the transition from the lead to the central region. 

The retarded single-scale propagator $\tilde S^R$ is
\begin{equation}
	\tilde S^R(\omega) = \left(\tilde G \tilde G_0^{-1} \partial_\Lambda \tilde G_0 \tilde G_0^{-1} \tilde G \right)^R = -\frac{i}{2} \tilde G^R_\Lambda \cdot \tilde G^R_\Lambda,
\end{equation}
where we omit the site and spin labels.

After the integration over the leads' degrees of freedom has been performed, the Green's function projected onto the central part acquires an additional self-energy term
\begin{equation}
	G^{R(\sigma)}_0(\omega) = \frac{1}{\omega^{(\sigma)} - \mathcal H^{(\sigma)} - {\Sigma^{(\sigma)}_{\rm lead}}(\omega, \Lambda) + i\Lambda/2},
\end{equation}
where $\omega^{(\sigma)} = \omega + \tfrac \sigma 2 B$ and
\begin{align}
\label{eq:Sigma_lead_flow}
	{\Sigma^{(\sigma)}_{\rm lead}}_{ij} (\omega, \Lambda) &= \frac{1}{2}\left(\omega^{(\sigma)} +i \tfrac{\Lambda}{2}-i\sqrt{4\tau^2-\left(\omega^{(\sigma)}+ 
i \tfrac{\Lambda}{2}\right)^2} \right) \nonumber \\
	                                &\times (\delta_{i,-N'} \delta_{j,-N'} + \delta_{i,N'} \delta_{j,N'}).
\end{align}
This self-energy is also reflected in the projected single-scale
propagator, which now takes the form
\begin{align}
	S^R(\omega) &= \left( G G_0^{-1} \partial_\Lambda G_0 G_0^{-1} G \right)^R \nonumber \\
	            &= G^R_\Lambda \cdot \left(-\frac{i}{2} + \partial_\Lambda \Sigma_{\rm lead} (\omega, \Lambda) \right) \cdot G^R_\Lambda.
\end{align}

For $\Lambda \rightarrow \infty$ the model is exactly solvable and the irreducible part of the full vertex is simply the bare vertex \cite{Jakobs2010-2}. Since we only consider equilibrium situations in this paper and the flow parameter respects fluctuation-dissipation theorems, the Keldysh Green's function $G^K$ [and single scale $S^K$] is determined simply via the fluctuation-dissipation theorem 
\begin{equation}
	G^K = (1-2 n_F)(G^R-G^A),~S^K = (1-2 n_F)(S^R-S^A).
\end{equation}

\subsection{B. The Vertex}
The vertex is assumed to consist \textit{only} of a two-particle contribution. This contribution is approximated by a structure compatible with a decomposition into three channels (with only static and local interchannel feedback). This approximation yields a consistent set of flow equations. We
use the following parametrization:

We decompose the $2$-particle vertex into three channels, according to
\begin{equation}
	\gamma(\omega_1', \omega_2'; \omega_1, \omega_2) \approx \bar v + \varphi^P(\omega_1 + \omega_2) + \varphi^X(\omega_2 - \omega_1') + \varphi^D(\omega_2 - \omega_2'),
\end{equation}
where we have suppressed all indices other than frequency, and primed quantities denote outgoing legs. $\bar v$ denotes the bare vertex.
The Keldsh structure is arranged according to the convention
\begin{equation}
	\gamma^{\alpha \beta | \gamma \delta} = 
	\begin{pmatrix}
		(qq|qq) & (qq|cq) & (qq|qc) & (qq|cc) \\
		(cq|qq) & (cq|cq) & (cq|qc) & (cq|cc) \\
		(qc|qq) & (qc|cq) & (qc|qc) & (qc|cc) \\
		(cc|qq) & (cc|cq) & (cc|qc) & (cc|cc) 
	\end{pmatrix}.
\end{equation}
The channels are labelled as (the Keldysh structure corresponds to Eqs.~(A8,A11,A17) of Ref.~\cite{Jakobs2010}, while the spatial structure is that of Eq.~(S48) of Ref.~\cite{Bauer2013})
	\begin{equation}
		(\varphi^P)_{(\sigma \bar \sigma | \sigma \bar \sigma), (ii|jj)}(\Pi) = 
		\begin{pmatrix}
		0           & a^{P*}_{ji} & a^{P*}_{ji} & 0          \\
		a^{P}_{ij}  & b^P_{ij}    & b^P_{ij}    & a^P_{ij}   \\
		a^{P}_{ij}  & b^P_{ij}    & b^P_{ij}    & a^P_{ij}   \\
		0           & a^{P*}_{ji} & a^{P*}_{ji} & 0
		\end{pmatrix}^{(\sigma \bar \sigma)} (\Pi),
	\end{equation}
	\begin{equation}
		(\varphi^X)_{(\sigma \bar \sigma | \sigma \bar \sigma), (ji|ij)}(X) = 
		\begin{pmatrix}
		0           & a^{X*}_{ji} & a^{X}_{ij}  & b^X_{ij}   \\
		a^{X}_{ij}  & b^X_{ij}    & 0           & a^{X*}_{ji}\\
		a^{X*}_{ji} & 0           & b^X_{ij}    & a^X_{ij}   \\
		b^X_{ij}    & a^{X}_{ij}  & a^{X*}_{ji} & 0           
		\end{pmatrix}^{(\sigma \bar \sigma)} (X),
	\end{equation}
	\begin{equation}
		(\varphi^D)_{(\sigma \sigma | \sigma \sigma), (ij|ij)}(\Delta) = 
		\begin{pmatrix}
		0           & a^{D}_{ij}  & a^{D*}_{ji} & b^D_{ij}   \\
		a^{D}_{ij}  & 0           & b^D_{ij}    & a^{D*}_{ii}\\
		a^{D*}_{ji} & b^D_{ij}    & 0           & a^D_{ij}   \\
		b^D_{ij}    & a^{D*}_{ji} & a^{D}_{ij}  & 0           
		\end{pmatrix}^{(\sigma \sigma)} (\Delta).
	\end{equation}
Each channel is labelled by only two spatial indices and one frequency. Conceptually, it can be thought of as the propagator of a Hubbard-Stratonovitch particle of the corresponding channel with retarded ($a^P$, $a^D$, and $a^{X*}$) and Keldysh ($b^P$, $b^{D}$, and $b^{X}$) components. From this point of view it is not surprising that in equilibrium the channels satisfy the fluctuation-dissipation theorems (c.f. Eqs.~(A10,A13,A19) of Ref.~\cite{Jakobs2010}:
\begin{subequations}
\begin{align}
	{b^P}^{(\sigma \bar \sigma)}_{(ij)} (\Pi)    & =   2 i \coth\left[\beta \left( \frac \Pi 2 - \mu \right)\right] {\rm Im} \, {a^{P}}^{(\sigma \bar \sigma)}_{(ij)} (\Pi) \\
	{b^X}^{(\sigma \bar \sigma)}_{(ij)} (X)      & = - 2 i \coth\left[\frac {\beta X} 2 \right]                     {\rm Im} \, {a^{X}}^{(\sigma \bar \sigma)}_{(ij)} (X) \\
	{b^D}^{(\sigma      \sigma)}_{(ij)} (\Delta) & =   2 i \coth\left[\frac {\beta \Delta} 2 \right]                {\rm Im} \, {a^{D}}^{(\sigma      \sigma)}_{(ij)} (\Delta)
\end{align}
\end{subequations}

\subsection{C. The Flow Equations}
When all vertices higher than the $2$-particle vertex are set to zero, the resulting truncated flow equations are (c.f. Eqs.~(27,28) of Ref.~\cite{Jakobs2010})
\begin{align}
	\frac{d}{d\Lambda} \Sigma^\Lambda_{1'1}       = 
& -\! \SumInt{2' 2}{}{4mm}  \frac{i}{2\pi} \gamma^\Lambda_{1' 2' 1 2} S^\Lambda_{2 2'} \nonumber \\
	\frac{d}{d\Lambda} \gamma^\Lambda_{1' 2' 1 2} = 
&+\! \SumInt{3' 4' 3 4}{}{6.5mm}  \!\!\! 
\frac{i}{2\pi} \gamma^\Lambda_{1' 2' 3 4} S^\Lambda_{3 3'} G^\Lambda_{4 4'} \gamma^\Lambda_{3' 4' 1 2} \nonumber
\\
&+\! \SumInt{3' 4' 3 4}{}{6.5mm} \!\!\! 
\frac{i}{2\pi} \gamma^\Lambda_{1' 4' 3 2} \left[S^\Lambda_{3 3'} G^\Lambda_{4 4'} + S^\Lambda_{4 4'} G^\Lambda_{3 3'} \right] \gamma^\Lambda_{3' 2' 1 4} \nonumber\\
&- \! \SumInt{3' 4' 3 4}{}{6.5mm} \!\!\! 
\frac{i}{2\pi} \gamma^\Lambda_{1' 3' 1 4} 
\left[S^\Lambda_{3 3'} G^\Lambda_{4 4'} + S^\Lambda_{4 4'} G^\Lambda_{3 3'} \right] 
\gamma^\Lambda_{4' 2' 3 2}.
\end{align}
Here, $1$, $1'$ etc. are multi-indices encompassing spin, site and
frequency. In the flow of the vertex, each summand corresponds to a
single channel. The vertex of each summand will be approximated by the
contribution of the corresponding channel for all frequencies and the
feedback of the other channels at a specific frequency ($2\mu$ for the
P-channel, $0$ for the X- and D-channels).  Inserting the channel
decomposition with the above notations into the flow equations, the
flow of the self-energy is given by [compare Eqs.~(B3,B4) of
Ref.~\cite{Jakobs2010}]:
\begin{widetext}
\begin{flalign}
	\partial_\Lambda \Sigma_{(kl)}^{q|c(\sigma)} (\omega) = -\frac{i}{2\pi} 
	                                                               \int d\omega' &\left[   S_{(lk)}  ^{c|c(\bar \sigma)} (\omega') {a^P}^{(\sigma \bar \sigma)}_{(kl)} (\omega + \omega')   \right.
	                                                                                      + S_{(kl)}  ^{c|c(\bar \sigma)} (\omega') {a^X}^{(\sigma \bar \sigma)}_{(lk)} (\omega' - \omega)  
	                                                                                      - S_{(kl)}  ^{c|c(     \sigma)} (\omega') {a^D}^{(\sigma \sigma)}_{(kl)}  (\omega - \omega')       \nonumber\\
	                                                                             &       + S_{(lk)}  ^{q|c(\bar \sigma)} (\omega') {b^P}^{(\sigma \bar \sigma)}_{(kl)} (\omega + \omega')  
	                                                                                      + S_{(kl)}  ^{c|q(\bar \sigma)} (\omega') {b^X}^{(\sigma \bar \sigma)}_{(lk)} (\omega' - \omega)  
	                                                                                      - S_{(kl)}  ^{c|q(     \sigma)} (\omega') {b^D}^{(\sigma \sigma)}_{(lk)} (\omega - \omega')     \nonumber\\
	                                                                             &       + S_{(lk)}    ^{c|c(\bar \sigma)} (\omega') U_k/2 \delta_{kl}                             
	                                                                                      + \left. \sum_m S_{(mm)}^{c|c(\sigma)} (\omega') {a^D}^{(\sigma \sigma)}_{(km)} (0) \delta_{kl} \right]          \nonumber\\
\end{flalign}
and
\begin{flalign}
	\partial_\Lambda \Sigma_{(kl)}^{q|q(\sigma)} (\omega) = -\frac{i}{2\pi} 
	                                                                        \int d\omega &'\left[                  S_{(kl)}  ^{c|q(\bar \sigma)}(\omega') {a^X}^{(\sigma \bar \sigma)}_{(lk)} (\omega' - \omega)     \right.
	                                                                                             -                 S_{(kl)}  ^{c|q(     \sigma)}(\omega') {a^D}^{(\sigma \sigma)}_{(kl)}      (\omega - \omega')            
	                                                                                             +                 S_{(lk)}  ^{q|c(\bar \sigma)}(\omega') {a^P}^{(\sigma \bar \sigma)}_{(kl)} (\omega + \omega')            \nonumber\\
	                                                                                     &       +                 S_{(lk)}  ^{c|q(\bar \sigma)}(\omega') {a^{P*}}^{(\sigma \bar \sigma)}_{(lk)} (\omega' + \omega)         
	                                                                                             +                 S_{(kl)}  ^{q|c(\bar \sigma)}(\omega') {a^{X*}}^{(\sigma \bar \sigma)}_{(kl)} (\omega' - \omega)         
	                                                                                             -                 S_{(kl)}  ^{q|c(     \sigma)}(\omega') {a^{D*}}^{(\sigma \sigma)}_{(lk)} (\omega - \omega')              \nonumber\\
	                                                                                     &       +                 S_{(lk)}  ^{c|c(\bar \sigma)}(\omega') {b^P}^{(\sigma \bar \sigma)}_{(kl)} (\omega + \omega')            
	                                                                                             +                 S_{(kl)}  ^{c|c(\bar \sigma)}(\omega') {b^X}^{(\sigma \bar \sigma)}_{(lk)} (\omega' - \omega)            
	                                                                                             -                 S_{(kl)}  ^{c|c(     \sigma)}(\omega') {b^D}^{(\sigma \sigma)}_{(kl)}      (\omega - \omega')            \nonumber\\
	                                                                                     &       + \left.\left( S_{(lk)}  ^{c|q(\bar \sigma)}(\omega') + S_{(lk)}  ^{q|c(\bar \sigma)} (\omega') \right) U_k/2 \delta_{kl}\right].\nonumber\\
\end{flalign}
The flow of the vertex contains two bubbles
\begin{align}
	I^{pp}_{ab|a'b'} &(\omega)^{(\sigma_1 \sigma_2)}_{(ij|kl)} = \frac{i}{2\pi} \int d\omega' \left[ G_{(i|k)}^{a|a'(\sigma_1)}(\omega/2 + \omega')S_{(j|l)}^{b|b'(\sigma_2)}(\omega/2 - \omega') +  S_{(i|k)}^{a|a'(\sigma_1)}(\omega/2 + \omega')G_{(j|l)}^{b|b'(\sigma_2)}(\omega/2 - \omega') \right],
\\
	I^{ph}_{ab|a'b'}& (\omega)^{(\sigma_1 \sigma_2)}_{(ij|kl)} = \frac{i}{2\pi} \int d\omega' \left[ G_{(i|k)}^{a|a'(\sigma_1)}(-\omega/2 + \omega')S_{(j|l)}^{b|b'(\sigma_2)}(\omega/2 + \omega') + S_{(i|k)}^{a|a'(\sigma_1)}(-\omega/2 + \omega')G_{(j|l)}^{b|b'(\sigma_2)}(\omega/2 + \omega') \right],
\end{align}
and is given by (compare Eqs.~(C3,C6,C9) of Ref.~\cite{Jakobs2010})
\begin{align}
\partial_\Lambda& (\varphi^P)^{qq|cq}_{(\sigma \bar \sigma \vert \sigma \bar \sigma)(ii \vert jj)} (\Pi) = \partial_\Lambda {a^{P*}}^{(\bar \sigma \sigma)}_{(ij)}(\Pi) \nonumber\\
		=& \sum_{km} \left(\tfrac 12 U_k \delta_{ki} + a^{P*}(\Pi)^{(\bar \sigma \sigma)}_{(ki)} + \tfrac{1}{2}{U^X}^{(\bar \sigma \sigma)}_{(ki)} \right) \left( I^{pp}_{cq|cc}(\Pi)^{(\sigma \bar \sigma|\sigma \bar \sigma)}_{(kk|mm)} + I^{pp}_{qc|cc}(\Pi)^{(\sigma \bar \sigma|\sigma \bar \sigma)}_{(kk|mm)}\right) \left( \tfrac 12 U_j \delta_{jm} + a^{P*}(\Pi)^{(\bar \sigma \sigma)}_{(jm)} + \tfrac{1}{2}{U^X}^{(\bar \sigma \sigma)}_{(jm)} \right)
\end{align}
\begin{align}
	\partial_\Lambda &(\varphi^X)^{qq|cq}_{(\sigma \bar \sigma \vert \sigma \bar \sigma)(ji|ij)} (X) = \partial_\Lambda a^{X*}(X)^{(\bar \sigma \sigma)}_{(ji)} \nonumber \\
	         =& \sum_{kl} \left(\tfrac{1}{2}U_j \delta_{jk} + \tfrac{1}{2}{U^P}^{(\bar \sigma \sigma)}_{(jk)} + a^{X*}(X)^{(\bar \sigma \sigma)}_{(jk)} \right) \left( I^{ph}_{qc|cc}(X)^{(\sigma \bar \sigma| \sigma \bar \sigma)}_{kl|lk} + I^{ph}_{cc|cq}(X)^{(\sigma \bar \sigma| \sigma \bar \sigma)}_{kl|lk} \right) \left( \tfrac{1}{2}U_i \delta_{il} + \tfrac{1}{2}{U^P}^{(\bar \sigma \sigma)}_{(li)} + a^{X*}(X)^{(\bar \sigma \sigma)}_{(li)} \right)
\end{align}
\begin{align}
	\partial_\Lambda &(\varphi^D)^{cq|qq}_{(\sigma \sigma)(ij|ij)} (\Delta) = \partial_\Lambda a^D(\Delta)^{(\sigma \sigma)}_{(ij)} \nonumber \\
	              =& -\sum_{kl} \left[ \left(-\tfrac 12 {W^D}^{(\sigma \sigma)}_{(ik)} + a^D(\Delta)^{(\sigma \sigma)}_{(ik)} \right) \left( I^{ph}_{qc|cc}(\Delta)^{(\sigma \sigma | \sigma \sigma)}_{(lk|kl)} + I^{ph}_{cc|cq}(\Delta)^{(\sigma \sigma | \sigma \sigma)}_{(lk|kl)} \right) \left(-\tfrac 12 {W^D}^{(\sigma \sigma)}_{(lj)} + a^D(\Delta)^{(\sigma \sigma)}_{(lj)} \right) \right.\nonumber \\
	               &+ \left. \left( \tfrac 12 {U_i} + \tfrac 12 {U^P}^{(\sigma \bar \sigma)}_{(ik)} + \tfrac 12 {U^X}^{(\sigma \bar \sigma)}_{(ik)} \right) \delta_{ik} \left( I^{ph}_{qc|cc}(\Delta)^{(\bar \sigma \bar \sigma | \bar \sigma \bar \sigma)}_{(lk|kl)} + I^{ph}_{cc|cq}(\Delta)^{(\bar \sigma \bar \sigma | \bar \sigma \bar \sigma)}_{(lk|kl)} \right) \delta_{jl} \left( \tfrac 12{U_j} + \tfrac 12 {U^P}^{(\bar \sigma \sigma)}_{(jl)} + \tfrac 12 {U^X}^{(\bar \sigma \sigma)}_{(jl)} \right) \right]
\end{align}
\begin{align}
	\partial_\Lambda &(\varphi^P)^{cq|cq}_{(\sigma \bar \sigma)(ii|jj)} (\Pi) = \partial_\Lambda b^P(\Pi)^{(\sigma \bar \sigma)}_{(ij)} \nonumber \\
		=& \sum_{km} \left[ \left(\tfrac{1}{2}U_i \delta_{ik} + a^P(\Pi)^{(\sigma \bar \sigma)}_{(ik)} + \tfrac{1}{2}{U^X}^{(\sigma \bar \sigma)}_{(ik)} \right) \left( I^{pp}_{cc|cc}(\Pi)^{(\sigma \bar \sigma|\sigma \bar \sigma)}_{(kk|mm)} + I^{pp}_{qq|cc}(\Pi)^{(\sigma \bar \sigma|\sigma \bar \sigma)}_{(kk|mm)} + I^{pp}_{cc|qq}(\Pi)^{(\sigma \bar \sigma|\sigma \bar \sigma)}_{(kk|mm)} \right) \right. \nonumber \\
		& \times \left(\tfrac{1}{2}U_j \delta_{jm} + a^{P*}(\Pi)^{(\sigma \bar \sigma)}_{(jm)} + \tfrac{1}{2}{U^X}^{(\sigma \bar \sigma)}_{(jm)} \right)\nonumber \\
		 &+ b^P(\Pi)^{(\sigma \bar \sigma)}_{(ik)} \left( I^{pp}_{qc|cc}(\Pi)^{(\sigma \bar \sigma|\sigma \bar \sigma)}_{(kk|mm)} + I^{pp}_{cq|cc}(\Pi)^{(\sigma \bar \sigma|\sigma \bar \sigma)}_{(kk|mm)} \right) \left(\tfrac{1}{2}U_j \delta_{jm} + a^{P*}(\Pi)^{(\sigma \bar \sigma)}_{(jm)} + \tfrac{1}{2}{U^X}^{(\sigma \bar \sigma)}_{(jm)} \right) \nonumber \\
		 &+ \left .\left(\tfrac{1}{2}U_i \delta_{ik} + a^P(\Pi)^{(\sigma \bar \sigma)}_{(ik)} + \tfrac{1}{2}{U^X}^{(\sigma \bar \sigma)}_{(ik)} \right) \left( I^{pp}_{cc|qc}(\Pi)^{(\sigma \bar \sigma|\sigma \bar \sigma)}_{(kk|mm)} + I^{pp}_{cc|cq}(\Pi)^{(\sigma \bar \sigma|\sigma \bar \sigma)}_{(kk|mm)} \right) b^P(\Pi)^{(\sigma \bar \sigma)}_{(mj)} \right]
\end{align}
\begin{align}
	 \partial_\Lambda &(\varphi^X)_{ji|ij}^{qq|cc} (X) = \partial_\Lambda b^X(X)^{(\sigma \bar \sigma)}_{(ij)} \nonumber \\
	         =& \sum_{kl} \left[ \left(\tfrac{1}{2}U_k \delta_{kj} + \tfrac{1}{2}{U^P}^{(\sigma \bar \sigma)}_{(kj)} + a^X(X)^{(\sigma \bar \sigma)}_{(kj)} \right) \left( I^{ph}_{cc|cc}(X)^{(\sigma \bar \sigma | \sigma \bar \sigma)}_{kl|lk} + I^{ph}_{qc|cq}(X)^{(\sigma \bar \sigma | \sigma \bar \sigma)}_{kl|lk} + I^{ph}_{cq|qc}(X)^{(\sigma \bar \sigma | \sigma \bar \sigma)}_{kl|lk} \right) \right. \nonumber \\
				 &\, \, \times  \left(\tfrac{1}{2}U_l \delta_{il} + \tfrac{1}{2}{U^P}^{(\sigma \bar \sigma)}_{(il)} + a^{X*}(X)^{\sigma \bar \sigma}_{(li)}\right) \nonumber \\ 
	          &+ b^X(X)^{(\sigma \bar \sigma)}_{(kj)} \left( I^{ph}_{qc|cc}(X)^{(\sigma \bar \sigma | \sigma \bar \sigma)}_{kl|lk} + I^{ph}_{cc|cq}(X)^{(\sigma \bar \sigma | \sigma \bar \sigma)}_{kl|lk} \right) \left(\tfrac{1}{2}U_l \delta _{il} + \tfrac{1}{2}{U^P}^{(\sigma \bar \sigma)}_{(li)} + a^{X*}(X)^{\sigma \bar \sigma}_{li}\right) \nonumber \\ 
	          &+ \left. \left(\tfrac{1}{2}U_j \delta_{jk} + \tfrac{1}{2}{U^P}^{(\sigma \bar \sigma)}_{(jk)} + a^X(X)^{(\sigma \bar \sigma)}_{(kj)}\right) \left( I^{ph}_{cq|cc}(X)^{(\sigma \bar \sigma | \sigma \bar \sigma)}_{kl|lk} + I^{ph}_{cc|qc}(X)^{(\sigma \bar \sigma | \sigma \bar \sigma)}_{kl|lk} \right) b^X(X)^{(\sigma \bar \sigma)}_{(il)} \right]
\end{align}
\begin{align}
	\partial_\Lambda &(\varphi^D)^{cc|qq}_{(\sigma \sigma)(ij|ij)} (\Delta) = \partial_\Lambda b^D(\Delta)^{(\sigma \sigma)}_{(ij)}  \nonumber \\
	              =& -\sum_{kl} \left[ \left(-\tfrac 12 {W^D}^{\sigma \sigma}_{ik} + a^D(\Delta)^{(\sigma \sigma)}_{(ik)}\right) \cdot \left( I^{ph}_{cc|cc}(\Delta)^{(\sigma \sigma | \sigma \sigma)}_{(lk|kl)} + I^{ph}_{qc|cq}(\Delta)^{(\sigma \sigma | \sigma \sigma)}_{(lk|kl)} + I^{ph}_{cq|qc}(\Delta)^{(\sigma \sigma | \sigma \sigma)}_{(lk|kl)} \right) \left(-\tfrac 12 {W^D}^{(\sigma \sigma)}_{(lj)} + a^{D*}(\Delta)^{(\sigma \sigma)}_{(jl)}\right) \right. \nonumber \\
	               &+ \left(-\tfrac 12 {W^D}^{(\sigma \sigma)}_{(ik)} + a^D(\Delta)^{(\sigma \sigma)}_{(ik)}\right) \left( I^{ph}_{qc|cc}(\Delta)^{(\sigma \sigma | \sigma \sigma)}_{(lk|kl)} + I^{ph}_{cc|cq}(\Delta)^{(\sigma \sigma | \sigma \sigma)}_{(lk|kl)} \right) b^D(\Delta)^{(\sigma \sigma)}_{(lj)} \nonumber \\
	               &+ b^D(\Delta)^{(\sigma \sigma)}_{(ik)} \left( I^{ph}_{cq|cc}(\Delta)^{(\sigma \sigma | \sigma \sigma)}_{(lk|kl)} + I^{ph}_{cc|qc}(\Delta)^{(\sigma \sigma | \sigma \sigma)}_{(lk|kl)} \right) \left(-\tfrac 12 {W^D}^{(\sigma \sigma)}_{(lj)} + a^{D*}(\Delta)^{(\sigma \sigma)}_{(jl)}\right) \nonumber \\
	               &+ \left. \left(\tfrac 12 {U_i} \delta_{ik} + \tfrac 12 {U^P}^{\sigma \bar \sigma}_{(ik)} + \tfrac 12 {U^X}^{\sigma \bar \sigma}_{(ik)} \right) \left( I^{ph}_{cc|cc}(\Delta)^{(\bar \sigma \bar \sigma | \bar \sigma \bar \sigma)}_{(lk|kl)} + I^{ph}_{qc|cq}(\Delta)^{(\bar \sigma \bar \sigma | \bar \sigma \bar \sigma)}_{(lk|kl)} + I^{ph}_{cq|qc}(\Delta)^{(\bar \sigma \bar \sigma | \bar \sigma \bar \sigma)}_{(lk|kl)} \right)
\right.\nonumber \\ & \phantom{+} 
\left.  \times 
\left(\tfrac 12 {U_l} \delta_{lj} + \tfrac 12 {U^P}^{(\bar \sigma \sigma)}_{(lj)} + \tfrac 12 {U^X}^{(\bar \sigma \sigma)}_{(lj)} \right) \right]
\end{align}
\end{widetext}
The relative signs between the $X$- and the $D$-channel stem from the fact that they are related through exchange of two fermionic legs.

In equilibrium, we set 
\begin{align}
{U^P}^{(\sigma \bar \sigma)}_{ij}   &= 2 {\rm Re} \, a^P(2 \mu)^{(\sigma \bar \sigma)}_{(ij)} \delta_{ij}, \nonumber \\
{U^X}^{(\sigma \bar \sigma)}_{(ij)} &= 2 {\rm Re} \, a^X (0)^{(\sigma \bar \sigma)}_{(ij)} \delta_{ij}, \nonumber \\
{W^D}^{(\sigma \sigma)}_{(ij)}      &= 2 {\rm Re} \, a^D (0)^{(\sigma \sigma)}_{(ij)} \delta_{ij}.
\end{align} 
Note that in equilibrium, $a^P(2 \mu)$, $a^X(0)$, and $a^D(0)$ are all real matrices.

In order to fully specify the flow, it remains to fix the initial conditions at large but finite Lambda:
\begin{align}
	\Sigma_{ij} & = \delta_{ij} U_i/2, \\
	\phi^X      & = \phi^P = \phi^D = 0.
\end{align}

\section{S-III. The frequency structure of the spin-susceptibility}
\label{sec:SS_freq}
\begin{figure}
\includegraphics[width=\columnwidth]{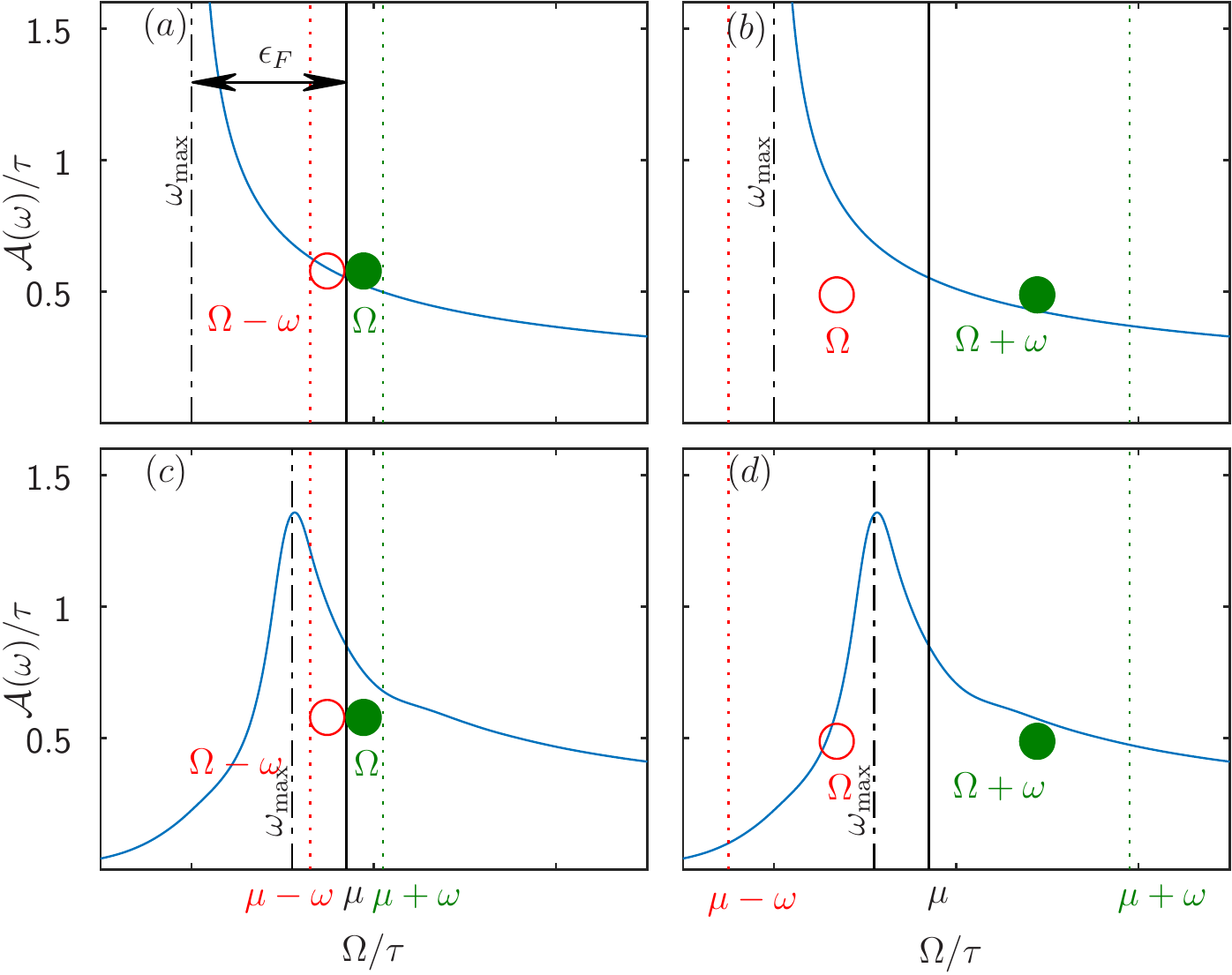}
\caption{(a,b) LDOS of a non-interacting homogenous system and (c,d) LDOS at the central site of an interacting QPC in the open regime. The filled (empty) circles show electrons (holes) of an electron-hole pair contributing to the spin susceptibility Eq.~\eqref{eq:nonint_local_spin_noise}, (a,c) for $\omega<\epsilon_F$ and (b,d) for $\omega>\epsilon_F$. Electron (or hole) energies lie between the chemical potential $\mu$ (solid black line) and $\mu+\omega$ (or $\mu-\omega$), indicated by the dotted green (or red) line. The frequency $\omega_{\rm max}$, at which the LDOS is maximal, is indicated by the black dashed-dotted line.}
\label{fig:eh-pairs}
\end{figure}
In this section, we substantiate the claim of the main text that the characteristic frequency $\omega_{\rm spin}$ of spin fluctuations is given by the distance between the chemical potential, $\mu$, and the lower effective band edge, $\omega_{\rm max}$.
To do so, we consider the local non-interacting spin-susceptibility, defined in Eq.~\eqref{eq:spin-noise_definition}, which at zero temperature can be written as
\begin{equation}
\label{eq:nonint_local_spin_noise}
	\chi^{U=0}_{ii} (\omega) = 2 \pi^2 \int_{\mu}^{\mu+\omega} d\Omega \mathcal A_i(\Omega) \mathcal A_i(\Omega-\omega).
\end{equation}
Let us begin by analyzing its properties for a homogenous tight-binding model with hopping $\tau$ and Fermi energy $\epsilon_F$ close to the lower band edge, i.e.~$\epsilon_F =\mu + 2 \tau \ll D, \omega \ll D$, where $D=4 \tau$ is the band width. This choice of $\epsilon_F$ most closely resembles the situation in the center of a QPC.
$\mathcal A_i (\Omega)$ is zero for frequencies below the band edge, shows a divergence at the band edge and subsequently decreases monotonically with increasing frequencies [Fig.~\ref{fig:eh-pairs} (a,b)].
$\chi^{U=0}_{ii}$ essentially counts the number of available electron-hole excitations, where the electrons have an energy of $\Omega \in [\mu, \mu+\omega]$ and the holes an energy $\Omega-\omega \in [\mu-\omega, \mu]$ [Fig.~\ref{fig:eh-pairs} (a,b)].

Consider $\omega < \epsilon_F$ [Fig.~\ref{fig:eh-pairs}(a)]. Then
\begin{align}
\label{eq:spin_noise_estimate1}
	\partial_\omega \chi^{U=0}_{ii} (\omega) = - 2 \pi^2 &\int_{\mu}^{\mu+\omega} d\Omega \mathcal A_i(\Omega) \mathcal A_i'(\Omega-\omega) \nonumber \\
	                                                    &+ 2 \pi^2 \mathcal A_i(\mu+\omega) \mathcal A_i(\mu) > 0.
\end{align}
Here, the prime denotes a derivative.
Thus $\chi^{U=0}_{ii} (\omega)$ is a monotonically increasing function for $\omega < \epsilon_F$.
This can be understood intuitively by considering the effects of an infinitesimal increase in $\omega$: The first term in Eq.~\eqref{eq:spin_noise_estimate1} describes how, if the \textit{electron} remains at energy $\Omega$, the weight of the hole at energy $\Omega-\omega$ increases [$\mathcal A_i'(\Omega-\omega)$]. The second term in Eq.~\eqref{eq:spin_noise_estimate1} describes the appearance of additional electron-hole pairs.

For $\epsilon_F < \omega$ [Fig.~\ref{fig:eh-pairs}(b)] Eq.~\eqref{eq:spin_noise_estimate1} is not useful, as the derivative of $\cal A$ is ill-defined at the band edge. We thus rewrite Eq.~\eqref{eq:nonint_local_spin_noise} as
\begin{equation}
\label{eq:nonint_local_spin_noise2}
	\chi^{U=0}_{ii} (\omega) = 2 \pi^2 \int_{\mu-\epsilon_F}^{\mu} d\Omega \mathcal A_i(\Omega+\omega) \mathcal A_i(\Omega),
\end{equation}
where we have used the fact that $\cal A$ vanishes for arguments below the band edge to restrict the range of integration.
Using Eq.~\eqref{eq:nonint_local_spin_noise2} we obtain
\begin{flalign}
\label{eq:spin_noise_estimate2}
	\partial_\omega \chi^{U=0}_{ii} (\omega) = 2 \pi^2 \int_{\mu-\epsilon_F}^\mu d\Omega \mathcal A'_i(\Omega+\omega) \mathcal A_i(\Omega) < 0.
\end{flalign}
For $\epsilon_F < \omega$, $\chi^{U=0}_{ii}(\omega)$ is thus monotonically decreasing. 
This can again be understood intuitively by considering the effects of an infinitesimal increase in $\omega$: consider an electron-hole pair with fixed \textit{hole} energy $\Omega$. The weight of the electron states near $\Omega+\omega$ [described by $\mathcal A_i'(\Omega+\omega)$] diminishes, reducing the spin susceptibility.

The above analysis and Eqs.~\eqref{eq:spin_noise_estimate1} and \eqref{eq:spin_noise_estimate2}, together, lead to the following important conclusion:
For the homogenous system considered so far, $\chi^{U=0}_{ii} (\omega)$ exhibits a local maximum at an energy, $\omega_{\rm spin}$, that corresponds to the Fermi energy, i.e.~to the distance between the chemical potential $\mu$ and the lower band edge $\omega_{\rm max}$, $\omega_{\rm  spin} = \mu - \omega_{\rm max}$.

We now switch to a QPC geometry in the presence of interactions. The inhomogenuity of the QPC potential changes the divergence of the bare LDOS at the band bottom into a broadened peak, but leaves the other features of the LDOS qualitatively unchanged [compare Fig.~\ref{fig:eh-pairs}(a) and (c) or (b) and (d)]. Within a Fermi liquid picture, where all of the above arguments still apply, albeit with renormalized parameters, we thus expect in the interacting QPC that $\omega_{\rm spin} \simeq \mu-\omega_{\rm max}$, where both $\omega_{\rm spin}$ and $\omega_{\rm max}$ are renormalized quantities.

\section{S-IV. Implementational Details}
The central region consists of $N=61$ sites. We use $\sim 1500$ frequencies to sample the real frequency axis. 
One third of the frequencies is sampled exponentially in the region $\vert \omega \vert > 4 \tau$, the rest is sampled homogeneously in the region $\omega \in [-4\tau, 4 \tau]$. An additional $~100$ frequencies are included in windows of size $4T$ around $\mu$ and $2 \mu$.
In order to numerically perform the integrals, it is useful to map the real axis to a finite region. We thus represent $\omega \in {\mathbbm R}$ in terms of the variable $\tilde y = y/\tau \in (-7, 7)$ via
\begin{equation}
\omega = 
\begin{cases}
	-2 \tau \frac{(\tilde y+6) (1+\Lambda)}{(\tilde y+6 )^2-1} -6 \tau,& \text{for } (\tilde y<-6) \\
	-2 \tau-\tau (\tilde y+2)^2/4,& \text{for } (-6 <\tilde y<-2) \\
	\tau \tilde y \sqrt{\frac{4}{\tilde y^2}-\frac{\tilde y^2-4^2}{4\tilde y^2}},& \text{for } (-2 < \tilde y<2) \\
	2 \tau+\tau (\tilde y-2)^2/4,& \text{for } (2 < \tilde y<6) \\
	-2 \tau \frac{(\tilde y-6)(1+\Lambda)}{(\tilde y-6)^2-1}+6 \tau,& \text{for } (6 < \tilde y).
\end{cases}
\end{equation}
The structure of this substitution is chosen such that the van Hove divergence at the band edges is trivially integrated (i.e.~the integral $\int d\omega (\omega \pm 2\tau)^{-1/2}$ is mapped to the integral ${\rm const.} \times \int d\tilde y 1$ for $\omega$ close to the band edges), while the large-frequency region is scaled with the flow parameter $\Lambda$ and substituted such that the integral $\int d\omega \omega^{-2}$ (the most diverging integral that occurs) is mapped to the integral ${\rm const.} \times \int d\tilde y 1$ for $\omega \gg \Lambda$. For convenience, $y=\pm 2 \tau, \pm 6 \tau$ is mapped to $\omega = \pm 2 \tau, \pm 6\tau$.
Continuous frequency information is obtained by linearly interpolating in $y$-space.

The flow equation is solved with a $6$th-order Runge-Kutta ODE solver with adaptive step size, while the integrals over internal frequencies are computed using Patterson sets.
The integrals over internal frequencies are split into multiple intervals, such that a strong dependence on the internal frequency occurs near the integration boundaries, as the sampling is more dense there.
The boundaries are determined by either the unsubstituted frequency of a Green's function or single scale propagator taking the value $\pm 2 \tau$, $\mu$, $\mu \pm 10 T$, $\pm (-2 \tau + \Vg)$ or $\pm 2\tau \pm \Lambda$, or by the argument of the P-channel (X-channel, D-channel) taking the value $2 \mu$ ($0$).
The flow parameter used is not $\Lambda$, but rather $u := \log \left( \frac{\Lambda}{1+\Lambda} \right)$. This improves the dynamic choice of step size within the ODE-solver. The flow starts at $\Lambda \approx 10^5$ and goes down to $\Lambda \approx 10^{-9}$.
To minimize runtime, the Green's function and single scale propagator are computed at $\sim 30000$ frequencies, and a linear interpolation in $y$-space is used when either of them is required in an integrand.
In equilibrium, the matrices appearing are symmetric under an exchange of sites. Further, the model considered here has a left-right parity symmetry. Both symmetries are exploited by using symmetric matrices to store the self-energy and the vertex, and by using a parity basis in the computation of the Green's function and the single scale propagator.

\section{S-V. DMRG calculations}
\begin{figure}
\includegraphics[width=\columnwidth]{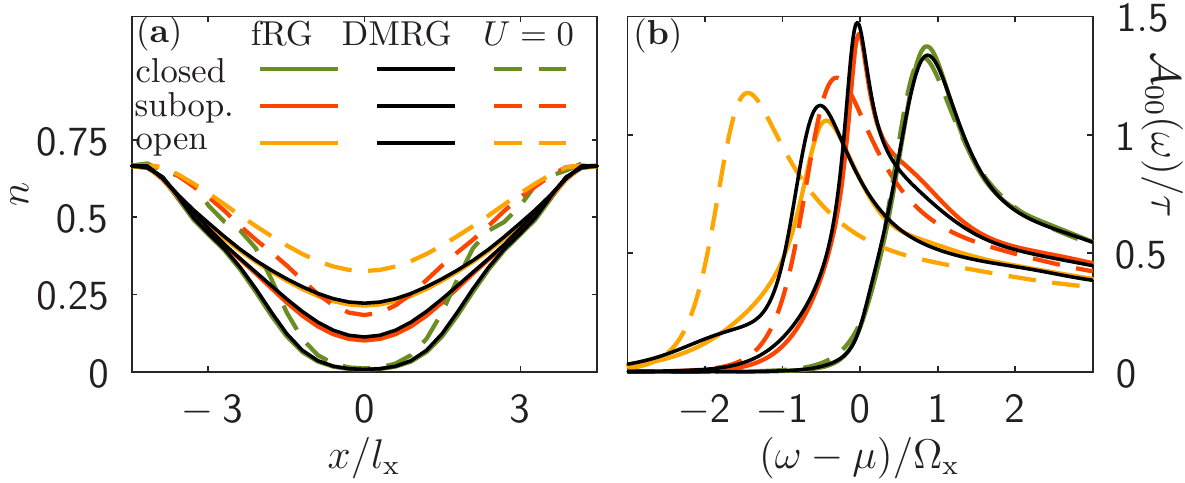}
\caption{Comparison between Keldysh-fRG and DMRG results. (a) The local density as a function of position and (b) the LDOS as a function of frequency of a closed (green), subopen (red), and open (orange) QPC, computed without interactions (dashed lines), and with interactions (solid lines), using Keldysh-fRG (colored) and DMRG (black), respectively.}
\label{fig:compare_DMRG}
\end{figure}
The results in the main text are obtained using Keldysh-fRG, which is
based on a perturbative ansatz. To verify the validity of the fRG data, we also employ
density-matrix-renormalization-group (DMRG) calculations
\cite{White1992,*White1993,*Schollwoeck2005}.  DMRG represents one of
the most powerful quasi-exact numerical method for describing
one-dimensional quantum many-body systems regarding their static
ground-state, dynamic, as well as thermodynamic properties. In
particular, DMRG can treat fermionic systems with arbitrary
interaction strength due to its non-perturbative character.
Specifically, we have used DMRG to compute the local density $n$ [Fig.~\ref{fig:compare_DMRG}(a)] and the LDOS [Fig.~\ref{fig:compare_DMRG}(b)], obtaining good qualitative agreement between our DMRG (black) and Keldysh-fRG (colored) results.

Below, we first elaborate on some peculiarities of our
matrix-product-state (MPS) implementation \cite{Schollwoeck2011},
which could be of interest to practitioners, and then discuss the choice of model parameters used for this comparison.

\subsection{A. DMRG details}

The QPC model in Eq.~\eqref{eq:define_H} poses a particular challenge to DMRG since, in contrast to fRG, it is not possible to incorporate the non-interacting leads to the left and right of the interacting region by an additional term in the self-energy. Instead, a finite-size chain representation of both leads is necessary as a prerequisite to make the model accessible for DMRG. The simplest ansatz is to replace the semi-infinite leads by a finite-length tight-binding chain with open boundary conditions (OBC). However, this setup is not practicable as it requires to go to very large system sizes in order to avoid strong finite-size artefacts in physical properties in the interacting part of the QPC. Instead, we here employ the concept of smooth boundary conditions (SBC) \cite{Vekic_PRL_1993,Vekic_PRB_1996}, which enable us to minimize finite-size effects in the interacting region of the QPC. Implementing SBC, the parameters of the non-interacting tight-binding chains are smoothly decreased to zero towards both ends of the chain to avoid having a sharp and rigid boundary as in the OBC setup. Thus for the interacting region of the QPC, the system's size is no longer fully determinable. SBC enable us to mimic very large leads with only $\mathcal{O}(10)$ sites. 

In practice, we scale the Hamiltonian parameters in the non-interacting regions (which we label symmetrically by $I=1, \dots, N_L$ for both the left and right lead; $I=1$ corresponds to the left- or rightmost boundary, $I=N_L$ to the lead sites closest to the central region) with a smoothing function $f_I$ such that $\tau_I=\tfrac \tau 2 (f_I + f_{I+1})$ and $\mu_I= \mu f_I$. Following Ref.~\cite{Vekic_PRB_1996}, we choose $f_I = y (1-I/[N_L+1]])$, and the smoothing function $y(x) = \tfrac 12 \big( 1- \tanh \frac{x-1/2}{x(1-x)}\big)$ for $0<x<1$, which interpolates between $1$ at the edge of the central region and $0$ and the boundary. 

In this setup, we first determine the ground state of the QPC using standard DMRG formulated in terms of MPS. The LDOS ${\cal A}_i (\omega) = -\frac 1\pi {\rm Im} G^{R}_{ii} (\omega)$ is then determined using time-dependent DMRG \cite{Vidal04,*Daley04,*WhiteFeiguin04}. To this end, we carry out two independent tDMRG runs to determine the retarded correlator in the time domain, ${G^{R}_{ii} (t) = -i[\langle c_i^\dagger(t) c_i \rangle+ \langle c_i(t) c_i^\dagger \rangle^*]}$. The entanglement in the MPS increases linearly during the real-time evolution, thus the number of states $D$ kept in simulation needs to be continuously increased to kept the numerical error constant. This implies that the simulation is bound to some maximum time $T_{\rm max}$ at which the simulation is no longer numerically feasible. A finite-time cutoff typically introduces artificial oscillations in the Fourier transform, requiring some artificial broadening to obtain a smooth and positive definite LDOS. However, we can avoid incorporating a broadening function by extending $T_{\rm max}$ to much larger times by means of linear prediction \cite{White_PRB_2008_LinPre,Barthel_PRB_2009}. The extrapolation scheme is expected to work for the present model since the correlator $G^{R}_{ii} (t)$ decreases exponentially over time scales smaller than the inverse mean level spacing and larger than the lifetime of excitations in the central region.

We end this section with some technical notes.
All DMRG calculations in this work are performed with the QSpace tensor library of A. Weichselbaum \cite{weichselbaum12a}. We studied
a QPC with an interacting region consisting of $N=31$ sites and two non-interacting regions to the left and right containing $N_L=50$ sites each, yielding a total of $N^{\rm DMRG}_{\rm tot} = 131$ sites, whose parameters are tuned in terms of SBCs (see above). The DMRG ground-state calculation employs a two-site update keeping up to $D=1600$ states. Convergence was typically reached after 10 to 40 sweeps, 40 being required particularly for an almost closed QPC, where the low particle density slows down convergence and the algorithm can get stuck in local minima during early iterations. In the tDMRG simulations we use a second-order Trotter-Suzuki decomposition with a time step $\Delta t=0.05/\tau$ and adapt the number of states in the MPS dynamically by truncating all singular values smaller than $\epsilon_{\rm SVD} = 5 \cdot 10^{-5}$. We stop the simulation when the number of kept states in the MPS exceeds $D=4000$. In this setting, we typically reach time scales $T_{\rm max} \cdot \tau = 60 - 65$ before applying linear prediction. 

\subsection{B. Choice of model parameters}

Since DMRG solves a finite system, we need a way to estimate the 'optimal' system size:
We extract the LDOS as a Fourier-transform of the real time Green's function, computed by DMRG. 
However, the resulting LDOS is only reliable if the Green's function is evolved up to time scales of the order of the traversal time $t_{\rm trav}$, as at shorter times the low-energy quasi-particles have yet to leave the central region. 
This means that the system size must be chosen sufficiently large, such that the reflection time $t_{\rm refl} \sim N^{\rm DMRG}_{\rm tot}/(2 \tau)$ (the time until the first quasi-particles reflected at the boundary return to the center) is larger than the traversal time: $t_{\rm refl} \gtrsim t_{\rm trav}$.
For the setup of the main text this yields $N^{\rm DMRG}_{\rm tot} \gtrsim 500$.
Combined with the fact that we need to perform time-evolution up to the traversal time $t_{\rm trav} \approx 250 / \tau$, this would have required an unfeasible amount of resources in DMRG.

In order to reduce the traversal time, we shrink the system (i.e.~reduce $N$) and make the QPC potential steeper (i.e.~increase the curvature $\Omega_{\rm x}$):
If the curvature is larger, a larger interaction is necessary to observe the same physics, as the LDOS is smeared out more.
We have tried to compensate for this by choosing an appropriately larger interaction.
Comparing Figs.~\ref{fig:resonant_energy_structure}(a) and \ref{fig:compare_DMRG}(b), we see that the qualitative features of the fRG-LDOS are the same: There is a roughly constant energy-shift of the LDOS in the open region, in the sub-open region the LDOS peak is sharpened (the effective potential is flatter) and pinned to the chemical potential, while the LDOS in the closed region is almost unaffected by interactions.
Since the new parameters yield results that exhibit the same qualitative features as those shown in the main text, we consider them a reasonable proxy for a direct comparison between DMRG and Keldysh-fRG.

To be specific, the set of parameters used for this comparison is:
$N^{\rm DMRG}=31$, $U^{\rm DMRG}=0.94 \tau$, $V_c^{\rm DMRG} = \{-1.69, -0.56, 0.56\} \Omega_{\rm x}^{\rm DMRG}$, $\mu^{\rm DMRG} = - \tau$, and $\Omega_{\rm x}^{\rm DMRG} \approx 0.9 \tau$.
Since $\Omega_{\rm x}$ is $3$ times larger than in the main text, the traversal time should be reduced by a factor of roughly $3$. 
We find $t_{\rm trav} \approx 70 / \tau$, and thus estimate $N_{\rm DMRG}^{\rm tot} \gtrsim 140$ (we use $N_{\rm DMRG}^{\rm tot} = 131$), which is still viable.

Finally, we remark that the choice of time $t_{\rm lp}$, after which linear prediction is applied, is a subtle issue:
The linear prediction method does not capture any physics that happens at time scales $t \gg t_{\rm lp}$ (this is an intended feature of the method, e.g.\ to mask finite-size effects).
However, this implies that for $t_{\rm trav} \gg t_{\rm lp}$ there may exist times at which linear prediction appears stable (i.e.~robust against variation of parameters used in linear prediction), while missing the finer details of the LDOS.
This happens in our system for times $t_{\rm lp} \sim 30 / \tau$, and is generically to be expected in a system with multiple time scales.
Once the largest time scale surviving the limit of infinite leads is reached (which in our case is $t_{\rm trav}$), and provided that time scale is still much shorter than the inverse level spacing, linear prediction appears to yield reasonable long-time results.

\end{document}